\documentclass[11pt,twocolumn,reprint,superscriptaddress,aps,nofootinbib,nolongbibliography,pre]{revtex4-2}

\usepackage{url,psfrag,graphicx}
\usepackage{amssymb}
\usepackage{bm}
\usepackage{caption}
\usepackage{comment}
\usepackage{dcolumn}
\usepackage{hyperref}
\usepackage{float,epsfig}
\usepackage{mathtools}
\usepackage{times}
\usepackage{multirow}
\usepackage[dvipsnames]{xcolor}
\usepackage{schemata} 

\begin{document}

\title{Large-scale kinetic roughening behavior of coffee-ring fronts}

\author{B. G. Barreales}
\affiliation{Departamento de F\'{\i}sica, Universidad de Extremadura, 06006 Badajoz, Spain}
\address{Instituto de Computaci\'on Cient\'{\i}fica Avanzada de Extremadura (ICCAEx), Universidad de Extremadura, 06006 Badajoz, Spain}
\author{J. J. Mel\'endez}
\affiliation{Departamento de F\'{\i}sica, Universidad de Extremadura, 06006 Badajoz, Spain}
\address{Instituto de Computaci\'on Cient\'{\i}fica Avanzada de Extremadura (ICCAEx), Universidad de Extremadura, 06006 Badajoz, Spain}
\author{R. Cuerno}
\affiliation{Departamento de Matem\'aticas and Grupo Interdisciplinar de Sistemas Complejos (GISC), Universidad Carlos III de Madrid, 28911 Legan\'es, Spain}
\author{J. J. Ruiz-Lorenzo}
\affiliation{Departamento de F\'{\i}sica, Universidad de Extremadura, 06006 Badajoz, Spain}
\address{Instituto de Computaci\'on Cient\'{\i}fica Avanzada de Extremadura (ICCAEx), Universidad de Extremadura, 06006 Badajoz, Spain}
\address{ Instituto de Biocomputaci\'on y F\'{\i}sica de
  Sistemas Complejos (BIFI), 50018 Zaragoza, Spain}
  
\date{\today}

\begin{abstract}

We have studied the kinetic roughening behavior of the fronts of coffee-ring aggregates via extensive numerical simulations of the off-lattice model considered for this context in [C.\ S.\ Dias {\it et al.}, Soft Matter {\bf 14}, 1903 (2018)]. This model describes ballistic aggregation of patchy colloids and depends on a parameter $r_\mathrm{AB}$ which controls the affinity of the two patches, A and B. Suitable boundary conditions allow us to elucidate a discontinuous pinning-depinning transition at $r_\mathrm{AB}=0$, with the front displaying intrinsic anomalous scaling, but with unusual exponent values $\alpha \simeq 1.2$, $\alpha_{\rm loc} \simeq 0.5$, $\beta\simeq 1$, and $z\simeq 1.2$. For $0<r_\mathrm{AB}\le 1$, comparison with simulations of standard off-lattice ballistic deposition indicates the occurrence of a morphological instability induced by the patch structure. As a result, we find that the asymptotic morphological behavior is dominated by macroscopic shapes. The intermediate time regime exhibits one-dimensional KPZ exponents for $r_\mathrm{AB}> 0.01$ and the system suffers a strong crossover dominated by the $r_\mathrm{AB}=0$ behavior for $r_\mathrm{AB}\le 0.01$. A detailed analysis of correlation functions shows that the aggregate fronts are always in the moving phase for $0<r_\mathrm{AB}\le 1$ and that their kinetic roughening behavior is intrinsically anomalous for $r_\mathrm{AB}\le 0.01$. 

\end{abstract} 

\maketitle 
\section{Introduction}
Non-equilibrium phenomena may feature complex non-trivial behaviors which often challenge intuition. These appear in a myriad of fields, including Biology \cite{Mori2011}, 
Engineering \cite{Albano1994}, Economy \cite{Arthur1999}, and, in general, whenever some ``equilibrium'' may be defined and somehow perturbed. In Physics, non-equilibrium is inherent to many different contexts, from fluid flow in porous media \cite{Hilfer1997}, to  the growth of thin films \cite{Ohring2002}, or very recently to driven-dissipative quantum systems \cite{Weimer2021}, to name just a few. Often it is closely related to the concepts of scaling and universality classes, the latter being understood as the set of systems which share a common scale-invariant limit. Both scaling properties and universality exist under equilibrium conditions too, of course, but it is out of equilibrium where they exhibit a richer phenomenology \cite{Tauber2014}.

The so-called coffee-ring effect is one of these non-trivial non-equilibrium behaviors, which arises in the context of Statistical Physics and impacts a number of industrial processes \cite{Laden1997}. Let us consider a drop of a liquid with suspended particles which dries up on a solid substrate. During the evaporation process, pinning of the contact line of the drop to the substrate causes the outward flow of liquid from the drop interior \cite{Deegan1997,Deegan2000}. The suspended solid particles are then dragged by capillary flow to the edges of the drop where they agglomerate in such a way that, after evaporation, they give rise to a characteristic ring-like stain. Thus, this familiar but complex effect results from the concomitance of diverse physical-chemical factors (capillarity, Marangoni flow), but also of geometrical ones. 

In this context, Yunker \textit{et al.}\ \cite{Yunker2011} showed that the structure of the ring-like pattern (but, interestingly, not the contact-line behavior of the capillary rates) could be sensitively altered by changes in the shape of the suspended particles. In particular, increasing the aspect ratio of the particles results into more complex, not necessarily ring-like, patterns, reaching the entire suppression of the coffee-ring effect for ellipsoidal particles of large enough eccentricity $\varepsilon$. The physical reason for this behavior, expected to hold even beyond the specific aqueous colloidal suspensions studied in Ref.\ \cite{Yunker2011,Dias2018}, is that elongated particles are dragged to the drop edge only until they reach the liquid-air interface, where they become subjected to long-range strong capillary attractions \cite{Loudet2005,Park2011}. As a result, loosely-packed particle clusters virtually arrested at the interface form and hinder further flow of particles; a ring-like pattern does not ensue in this case, but rather an homogeneous stain is observed. Unlike eccentric particles, spherical ones flow much more effectively to the drop edge, where they get concentrated and leave a ring after evaporation. Still, adding some eccentric particles to a suspension of spherical ones also suppresses the coffee-ring effect, provided that the diameter of the spheres is larger than the minor axis of the ellipsoids \cite{Yunker2011}.  
In a later work, Yunker \textit{et al.}\ characterized the space-time evolution of the front of the particle aggregate, by measuring its height $h$ and width or roughness $w$ as functions of time \cite{Yunker2013}. Irrespective of the value of the eccentricity, the front width increased in time following a power law $w(t) \sim t^\beta$, with $\beta$ being the so-called growth exponent characteristic of surface kinetic roughening \cite{Barabasi1995,Krug1997}. For spherical particles ($\varepsilon = 1$) $\beta \simeq 1/2$ was measured, consistent with a random deposition behavior \cite{Barabasi1995}. Increasing the eccentricity of the suspended particles to $\varepsilon = 1.2$ changes the growth exponent to $\beta \simeq 1/3$, characteristic of the one-dimensional (1D) Kardar-Parisi-Zhang (KPZ) universality class; further consistence with KPZ scaling was confirmed by the roughness exponent $\alpha \simeq 1/2$ (see Sec.\ \ref{sect:observables} below for precise definitions) and by the skewness and kurtosis of the distribution of height fluctuations, both of which are observed to take values consistent with those of the Tracy-Widom (TW) distribution exhibited by 1D-KPZ processes \cite{Kriecherbauer2010,Takeuchi2018}. Finally, for even more eccentric particles ($\varepsilon > 2.5$), the dynamic evolution of the front yields larger values $\alpha \simeq 0.61$, $\beta \simeq 0.68$, which the authors interpreted as consistent with the universality class of the KPZ equation with quenched disorder (QKPZ); see Refs.\ \cite{Barabasi1995,Galluccio1995,Stepanow1995,Tang1995,Leschhorn1996} and Appendix \ref{app:qkpz} for key facts on the QKPZ equation.

The latter fact is surprising since sources of quenched disorder cannot be unambiguously identified in the experiment \cite{Nicoli2013,Yunker2013reply}. Yunker \textit{et al.}\ argued that highly eccentric particles reaching the air-liquid interface at some sites become preferential locations for further deposition of particles by virtue of the aforementioned strong capillary attractions, in detriment of regions lacking particles. A non-homogeneous growth yielding QKPZ exponent values could be caused then by a colloidal ``Matthew effect'' \cite{Yunker2013}, in the sense that particle-rich regions become richer and particle-depleted ones remain poorly populated \cite{Merton1968}. However, as eloquently demonstrated e.g.\ by the supplemental video 3 of the experiments in Ref.\ \cite{Yunker2013}, the colloidal Matthew effect implies a dynamical instability for the front morphology in which quenched disorder plays no role.
This interpretation was tested by numerical simulations \cite{Nicoli2013} of the lattice growth model introduced in Ref.\ [\onlinecite{Yunker2013}] to describe the large $\varepsilon$ experiments, which showed that the ensuing instability indeed leads to large values of the scaling exponents, compatible with those measured at large eccentricities. More specifically, at large $\varepsilon$ the Matthew effect leads to anomalous kinetic roughening for the front \cite{Nicoli2013,Lopez1997,Ramasco2000,Cuerno2004}. In particular, the critical exponent values may not be universal, but rather depend on geometry and physical parameters. 

The origin and features of QKPZ scaling for the coffee-ring effect have attracted considerable interest. Oliveira and Aar\~ao Reis \cite{Oliveira2014} used a lattice  ballistic-deposition-like model \cite{Barabasi1995} based on the so-called RCA model \cite{Rodriguez-Perez2005}, in which particles fall towards the substrate by moving one position down vertically and $D$ positions horizontally. Thus, the parameter $D$ models the aspect ratio of the suspended particles; in particular, $D=0$ corresponds to ballistic deposition. Oliveira and Aar\~ao Reis obtained $\beta \simeq 0.33$ (i.e., the expected 1D-KPZ value) for $D=0$ and $\beta \simeq 0.68$ (compatible with 1D-QKPZ) for $D = 8$. While these extreme values agree with those measured in Ref.\ [\onlinecite{Yunker2013}], the continuous variation of $\beta$ with $D$ contrasts with the experimental observations. Reference [\onlinecite{Oliveira2014}] also reports the value of the dynamic exponent $z$ (see Sec.\ \ref{sect:observables}) characterizing the growth of the correlation length along the interface \cite{Barabasi1995,Krug1997}; specifically, $z \simeq 1.56$ for $D=0$, compatible with 1D-KPZ behavior ($z = 1.5$), but $z \simeq 2.56$ for $D = 8$, well above the 1D-QKPZ prediction ($z = 1)$. The authors concluded that the $\beta$ and $z$ exponents computed for large $D$ were caused by the ``columnar'' front growth, with no need to invoke any quenched disorder, in qualitative agreement with the analysis performed in Ref.\ [\onlinecite{Nicoli2013}].

\begin{figure}[!t]
    \centering
    \includegraphics[width=0.4\textwidth]{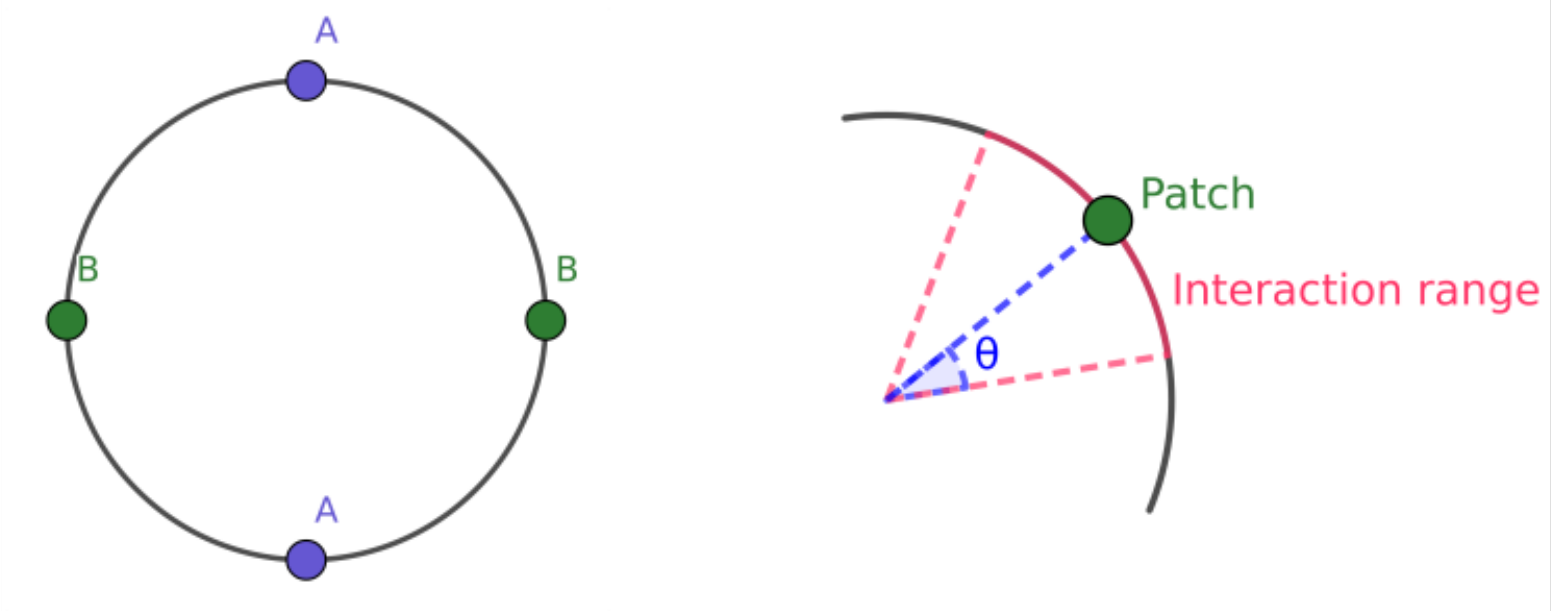}
    \caption{Left panel: Single patchy colloidal particle with four patches, as considered in the model of Ref.\ [\onlinecite{Dias2013}]. Right panel: Interaction range around a patch, described by $\theta$, see Sec.\ \ref{sect:TheModel} and Appendix \ref{app:details} for additional details.}
    \label{fig:patches}
\end{figure}

An alternative theoretical approach to describe the kinetic roughening behavior seen in the experiments of Ref.\ [\onlinecite{Yunker2013}] was later taken by Dias \textit{et al}.\ \cite{Dias2013,Dias2014,Dias2018}, based on the deposition of ``patchy'' colloids with weak and strong bonds. More specifically, an off-lattice model is considered in which circular particles with patches fall vertically onto a flat substrate. In order to model the anisotropy of the experimental ellipsoidal colloids using circular disks, the latter are assumed to have two patch pairs, namely A at the poles and B along the equator, see Fig.\ \ref{fig:patches}. The falling particles eventually aggregate to the substrate, or to already deposited particles, via patch-patch interactions ruled by three binding probabilities, $P_\mathrm{AA}$, $P_\mathrm{AB}$, and $P_\mathrm{BB}$, corresponding to the three possible patch-patch configurations. Dias \textit{et al}.\ found two different regimes depending on $r_\mathrm{AB} \equiv P_\mathrm{AB}/P_\mathrm{AA}$, with $P_\mathrm{AB} = P_\mathrm{BB}$ for simplicity. For large $r_\mathrm{AB}$ values $0.5 \le r_\mathrm{AB} \leq 1$ (i.e., small particle eccentricity), scaling exponents are consistent \cite{Dias2014,Dias2018} with 1D-KPZ behavior ($\beta = 1/3$ and $z = 3/2$ \cite{Barabasi1995,Krug1997}); for small $0.01 < r_\mathrm{AB} < 0.1$ (i.e., large particle eccentricity), these simulations obtain $\beta \simeq 0.63$ and $z \simeq 1$, which was ascribed to 1D-QKPZ scaling ($\beta = 0.63$ and $z$ = 1, \cite{Leschhorn1996}), crossover behavior being obtained for intermediate $0.1 < r_\mathrm{AB} < 0.5$. In general, the average front velocity is nonzero for any value of $r_{\rm AB}$ in the simulations of Refs.\ \cite{Dias2014, Dias2018}, at variance with the QKPZ equation. Indeed, this continuum model is well-known to display a pinning transition (see Appendix \ref{app:qkpz}) between a pinned phase, in which the average front velocity is zero, and a moving phase, in which it is non-zero. The transition is termed directed percolation depinning (DPD), as it is induced by the emergence of a directed percolation cluster of quenched disorder sites where front motion is arrested \cite{Barabasi1995,Galluccio1995,Tang1995,Leschhorn1996}. The scaling exponents measured in Refs.\ \cite{Dias2014, Dias2018} for
$0.01 \le r_\mathrm{AB} \le 0.1$
are those of the QKPZ equation right at the transition point, while the moving phase of the DPD transition features still larger exponents $\alpha_{\rm mp}=0.75$ and $\beta_{\rm mp}=0.74$ \cite{Makse1995,Amaral1995}. Hence, the peculiar conclusion in Refs.\ \cite{Dias2014, Dias2018} is that a pinning transition takes place for each value of $r_\mathrm{AB}$ in the finite interval $[0.01,0.1]$.

In this paper we perform a systematic study of the kinetic roughening behavior of the fronts produced by the off-lattice aggregation model of Dias \textit{et al}.\ in the full $r_\mathrm{AB} \in [0,1]$ parameter range, in which we consider longer evolution times and larger system sizes. Beyond the analysis of the surface roughness performed in Refs.\ [\onlinecite{Dias2014}] and [\onlinecite{Dias2018}], here we compute and analyze additional front correlation functions in real and reciprocal space. Our results show that a discontinuous pinning-depinning phase transition does exist at $r_\mathrm{AB} = 0$. The $r_\mathrm{AB}=0$ case (\rm{pinned front}) exhibits unusual critical exponents, while the long time dynamics for $r_\mathrm{AB}>0$ (moving front) is dominated by a morphological instability that induces large scaling exponent values and fronts dominated by macroscopic shapes at large time scales and system sizes, as well as intrinsic anomalous scaling for the smaller $r_\mathrm{AB}$. Besides, QKPZ scaling is an effective behavior seen for intermediate times and system sizes, and suitable values of $r_\mathrm{AB}$.

The paper is structured as follows: In Sec.~\ref{sect:TheModel} we present all remaining details of the model; in Sec.~\ref{sect:observables} we describe all the observables used in this paper. The results of our numerical simulations are presented in Sec.~\ref{sect:results}, which is followed by a discussion in Sec.\ \ref{sec:disc}, and by a summary and our conclusions in Sec.\ \ref{sect:conclusions}. Finally, four appendices complete the paper; as already noted, one of them provides background on the QKPZ equation, while three additional ones are devoted to technical issues.

\section{Model}
\label{sect:TheModel}

We consider the patchy colloids model developed in Refs.\ \cite{Dias2013,Dias2014}. As introduced in the previous section, two-dimensional circular colloids of radius $R$ fall down vertically at a random horizontal position, either to get adsorbed onto a one-dimensional planar substrate, or to bind to already present colloids. Each colloid contains two types of patches, namely two A-type patches at the poles and two B-type ones along the equator, see Fig.\ \ref{fig:patches}. If a deposited colloidal particle falls on top of a previous one, it binds to it with a certain probability that is assumed to be thermally activated, and modeled by three binding probabilities, $P_\mathrm{AA}$, $P_\mathrm{BB}$, and $P_\mathrm{AB}$, defined as Arrhenius-like functions of temperature $T$,
\begin{equation}
    P_i \propto e^{-E_a^{(i)} / k_\mathrm{B} T},
    \label{eq:arrhenius}
\end{equation}
where $k_\mathrm{B}$ is Boltzmann's constant and $E_a^{(i)}$ holds for the activation energy which characterizes the strength of each bond type ($i=$ AA, AB, and BB). Following Ref.\ \cite{Dias2014}, we will assume $P_\mathrm{AA}=1$ without loss of generality and $P_\mathrm{AB}=P_\mathrm{BB}$ for simplicity, and define the sticking coefficient $r_\mathrm{AB} = P_\mathrm{AB}/P_\mathrm{AA}$. The colloid-colloid interaction is limited to a region spanning an angle $\theta=\pi/6$ around each patch, as depicted in Fig.\ \ref{fig:patches}. Binding only occurs if the interaction ranges of the adjacent colloids overlap; in such a case, the newly aggregated colloidal particle reorients itself so that its patches get aligned with those of the pre-existing particle to which it attaches. Additional simulation details are given in Appendix \ref{app:details} \cite{Orozco2019}. According to these definitions, low $r_\mathrm{AB}$ values favor AA interactions, and the resulting aggregate morphology resembles that experimentally observed for ellipsoidal colloids; the $r_\mathrm{AB} = 0$ limit corresponds to the situation for which only AA unions are favorable and long 1D chains of particles form, see below. On the contrary, $r_\mathrm{AB} = 1$ mimics the behavior of circular (isotropic) particles, which is the condition for the formation of a coffee-ring aggregate, properly speaking \cite{Dias2018}. Note, however, that, due to the existence of the finite interaction range with angle $\theta$ and the related alignment step in the attachment process, the $r_\mathrm{AB} = 1$ case is not identical to the simple off-lattice ballistic deposition model, well known to belong to the KPZ universality class \cite{Meakin1998}. This difference will be seen below to play a relevant role.

In all cases, free boundary conditions (FBC) were chosen along the substrate direction. The reason for such a choice, instead of the common periodic boundary conditions (PBC) \cite{Barabasi1995}, is the morphological evolution of the system for $r_\mathrm{AB} = 0$. Indeed, the system grows indefinitely for $r_\mathrm{AB} = 0$ under PBC, which conflicts with the expected behavior that the system eventually reaches a steady state at finite time $t$, to be discussed below. For consistency, FBC were used for non-zero $r_\mathrm{AB}$ values as well. In Appendix~\ref{app:PBC} we provide a comparison between the results obtained for both types of boundary conditions. 

The uncertainties of all computed values have been calculated following the jackknife procedure \cite{Young2015,Efron1982}; see Appendix C in Ref.~\cite{Barreales2020} for additional details.

\section{Observables}
\label{sect:observables}

A two dimensional lattice with lateral length $L$ is divided into $L$ columns with equal widths $\Delta = 2R =1$. At each column $i$, the front $h(x_i,t)$ is defined as the $y-$coordinate of the highest particle in this column, even if only part of it is there; hereafter, the front position will be denoted as $h(x,t)$ for simplicity. The mean front is computed then as 
\begin{equation}
\overline{ h(t)}=\frac{1}{L}\sum_{i=1}^{L}h(x_i,t)\,,
\end{equation}
which defines the overline symbol, $\overline{ (\cdots)}$.

The front width, or front roughness $w(L,t)$, is defined as the standard deviation of the front values,
\begin{equation}
    w^2(L,t)=\left\langle \overline{[h(x,t)-\overline{h(t)}]^2} \right\rangle,
    \label{eq:w}
\end{equation}
where  $\langle (\cdots) \rangle$ denotes average over different realizations of the noise. Under the simplest kinetic roughening conditions, the roughness $w(L,t)$ is expected to satisfy the Family-Vicsek (FV) scaling relation \cite{Barabasi1995, Krug1997}
\begin{equation}
	\label{eq:wdef}
	w(L,t)=t^{\beta}f\left(t/L^z \right),
\end{equation}
where $\beta$ and $z$ are the growth and dynamic exponents, respectively, and the scaling function has two asymptotic limits. For $t \ll L^z$ one has $f(y) \sim \text{const}$, thus $w\sim t^{\beta}$ in such a limit. On the other hand, for $t \gg L^z$, it is $f(y) \sim y^{-\beta}$, so that $w = \text{const} \equiv w_{\mathrm{sat}}$, with $w_{\mathrm{sat}}$ being the value the roughness saturates into at steady state, which in turn scales with the lateral size of the system as $w_{\mathrm{sat}} \sim L^{\alpha}$; here, $\alpha$ is the so-called roughness exponent. Note that $\alpha$, $\beta$, and $z$ are related by $\alpha=\beta z$ \cite{Barabasi1995, Krug1997}. 

According to Eq.\ \eqref{eq:w}, the growth exponent $\beta$ characterizes the time-dependent dynamics of the roughening process. On the other hand, the $\alpha$ exponent is related with the fractal dimension of the front \cite{Barabasi1995}. Finally, the dynamic exponent $z$ quantifies the power-law increase of the lateral correlation length along the front \cite{Barabasi1995,Tauber2014},
\begin{equation}
    \label{eq:correlation_length}
    \xi(t) \sim t^{1/z} .
\end{equation}

Additional insights about the dynamic evolution will be provided by the height-difference correlation function $C_2(r,t)$, defined as
\begin{equation}
\label{eq:correlation_2}
\begin{split}
	C_2(r, t) & =\frac 1{L} \sum_x \left\langle [h(x+r, t)-h(x, t)]^2 \right\rangle \\
	& = 2\langle \overline{h(t)^2}\rangle-\frac 2{L}\sum_x \langle h(r+x, t)h(x, t) \rangle, 
\end{split}
\end{equation}
where the sum is over all $x$ values. Under FV kinetic roughening conditions, 
\begin{equation}
    C_2(r,t)=r^{2\alpha} g_{\mathrm{FV}}(r/\xi(t)) ,
    \label{eq:corr_lengthFV}
\end{equation}
where $g_{\mathrm{FV}}(u) \sim u^{-2\alpha}$ for $u\gg 1$ and $g_{\mathrm{FV}}(u)\sim {\rm const}$ for $u\ll1$ \cite{Barabasi1995,Krug1997}. In particular, $C_2(r,t)$ saturates (i.e., becomes $r$-independent) as $C_2(r,t) \sim \xi^{2\alpha}(t) $ for $r\gg \xi(t)$. 
This behavior is analogous to what is expected for the so-called local roughness, defined like the global roughness, Eq.\ \eqref{eq:wdef}, but only within a local box of size $r<L$ \cite{Barabasi1995,Krug1997}. Moreover, the height-difference correlation function allows one to describe the spatio-temporal evolution of the front and to evaluate the correlation length $\xi(t)$. Indeed, one may write
\begin{equation}
C_2(\xi_a(t),t)= a \, C_{2,p},    
    \label{eq:compute_corr_length}
\end{equation}
where $a$ is a constant taken arbitrarily as $a = 0.9$ in our case; the precise value of $a$ does not modify the scaling relation \cite{Barreales2020}. In this work, we define the correlation length at a given time $t$ as the distance along the front at which the $C_2$ function takes 90\% of its plateau-value, with the plateau-value being estimated as $C_{2,p}=\overline{C_2(r)}$ for $0.7L<r<0.9L$. 

There are cases in which the FV scaling Ansatz for the correlation function, Eq.\ \eqref{eq:corr_lengthFV}, needs to be generalized into \cite{Lopez1997,Ramasco2000,Cuerno2004}
\begin{equation}
    C_2(r,t)=r^{2\alpha} g(r/\xi(t)),
    \label{eq:corr_length}
\end{equation}
where $g(u) \sim u^{-2\alpha}$ for $u\gg 1$ and $g(u) \sim u^{-2(\alpha-\alpha_{\rm loc})}$ for $u\ll 1$. Now $\alpha_{\text{loc}}$ is a so-called local roughness exponent which characterizes the front fluctuations measured at distances smaller than the system size $L$. Under FV scaling, the two roughness exponents are equal \cite{Barabasi1995,Krug1997}, $\alpha=\alpha_{\text{loc}}$, so that $g(u)=g_{\mathrm{FV}}(u)$ and Eq.\ \eqref{eq:corr_length} coincides with Eq.\ \eqref{eq:corr_lengthFV}. However, there are cases in which $\alpha_{\text{loc}}\neq \alpha$, so that front fluctuations at small and large distances are characterized by two independent roughness exponents. If $\alpha_{\text{loc}}<1$, these systems are said to display intrinsic anomalous kinetic roughening \cite{Lopez1997,Ramasco2000,Cuerno2004}.

Anomalous scaling may be originated by different causes \cite{Cuerno2004} and can be also characterized efficiently \cite{Lopez1997} through the front structure factor $S(q,t)$, defined as
\begin{equation}
    S(q,t)=\langle |\mathcal{F}[\delta h(x,t)]|^2 \rangle ,
\end{equation}
where $\mathcal F$ denotes the space Fourier transform of the front fluctuations $\delta h(x,t)=h(x,t)-\overline{h(t)}$ and $q$ is the magnitude of an one-dimensional wave vector. In terms of the structure factor, the scaling relation reads
\begin{equation}
 S(q,t)=q^{-(2\alpha+1)}s(qt^{1/z}),
 \label{eq:scaling_s}
\end{equation}
where $s(u)\sim u^{2\alpha+1}$ for $u\ll 1$ and $s(u) \sim u^{2(\alpha-\alpha_{\rm loc})} $ for $u\gg 1$. Under Family-Vicsek scaling $\alpha=\alpha_{\text{loc}}$ and the scaling function reads $s(u)\sim {\rm const}$ for $u\gg 1$. Otherwise, the system exhibits intrinsic anomalous scaling if the structure factor scales as in Eq.~\eqref{eq:scaling_s}; in particular, this implies that, for $q\gg 1/t^{1/z}$, the structure factor scales with the wave-vector magnitude as $S(q,t)\sim q^{-(2\alpha_{\text{loc}}+1)}$.
 
\begin{figure}[!tb]
    \centering
    \includegraphics[width=0.45\textwidth]{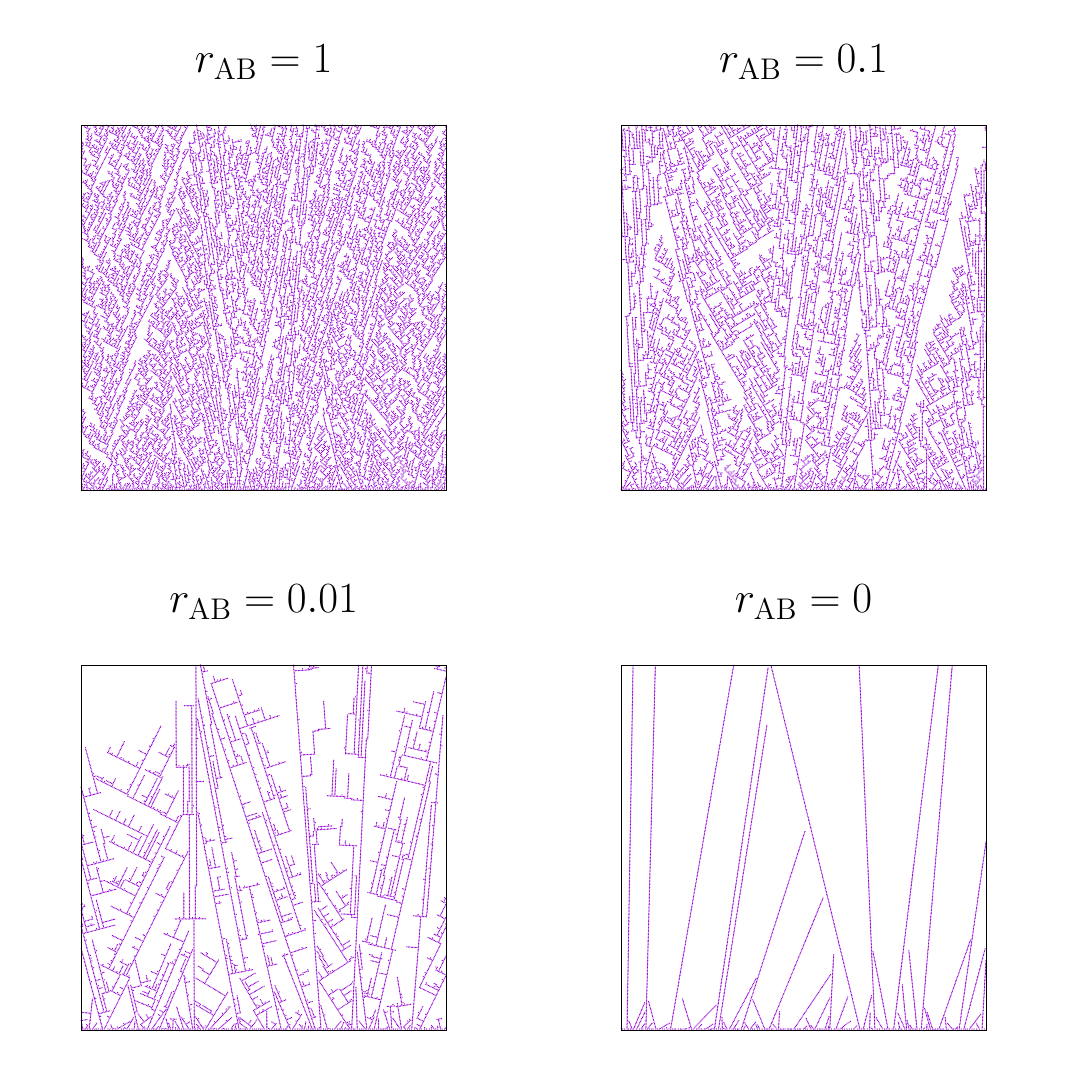}
    \caption{Morphologies of the colloidal aggregates calculated for values of $r_\mathrm{AB}$ (approximately, inverse colloid eccentricities) as indicated on each panel. Single colloidal particles appear as bullets. All the snapshots were computed for $L = 256$ for times ranging from the initial ones to final times for which the points of the morphologies fill the plots.}
    \label{fig:morf_prob}
\end{figure}

\section{Results}
\label{sect:results}

\subsection{Aggregate morphology}

The morphology of the colloidal aggregates changes with $r_\mathrm{AB}$, as shown in Fig.~\ref{fig:morf_prob} for $r_\mathrm{AB}=1, 0.1, 0.01$, and $0$. The overall appearance is quite similar for the highest $r_\mathrm{AB}$ values, featuring a tangled mess of branches oriented along random directions. In particular, the snapshots for $r_\mathrm{AB} = 1$ and $r_\mathrm{AB} = 0.1$ qualitatively resemble the experimental evidence of closely-packed agglomeration of nearly-spherical (i.e., $\varepsilon~\simeq$~1.0 - 1.5) colloids \cite{Dias2018}. These model morphologies also recall those of the simple off-lattice ballistic deposition model, whose kinetic roughening fronts are well known instances of KPZ scaling behavior \cite{Barabasi1995,Meakin1998}. The density of secondary branches, starting from pre-existing ones and not from the substrate, decreases as $r_\mathrm{AB}$ (and thence the AB binding probability) does, again in agreement with the experimental observation that loosely-packed aggregates form when the colloids are highly elongated. In the limiting $r_\mathrm{AB} = 0$ case, secondary branches are completely suppressed, and the front grows only along some directions, defined by the first deposited particles. 

\subsection{Front velocity}
\label{subsec:vel}

\begin{figure}[thb]
    \centering
    \includegraphics[width=0.45\textwidth]{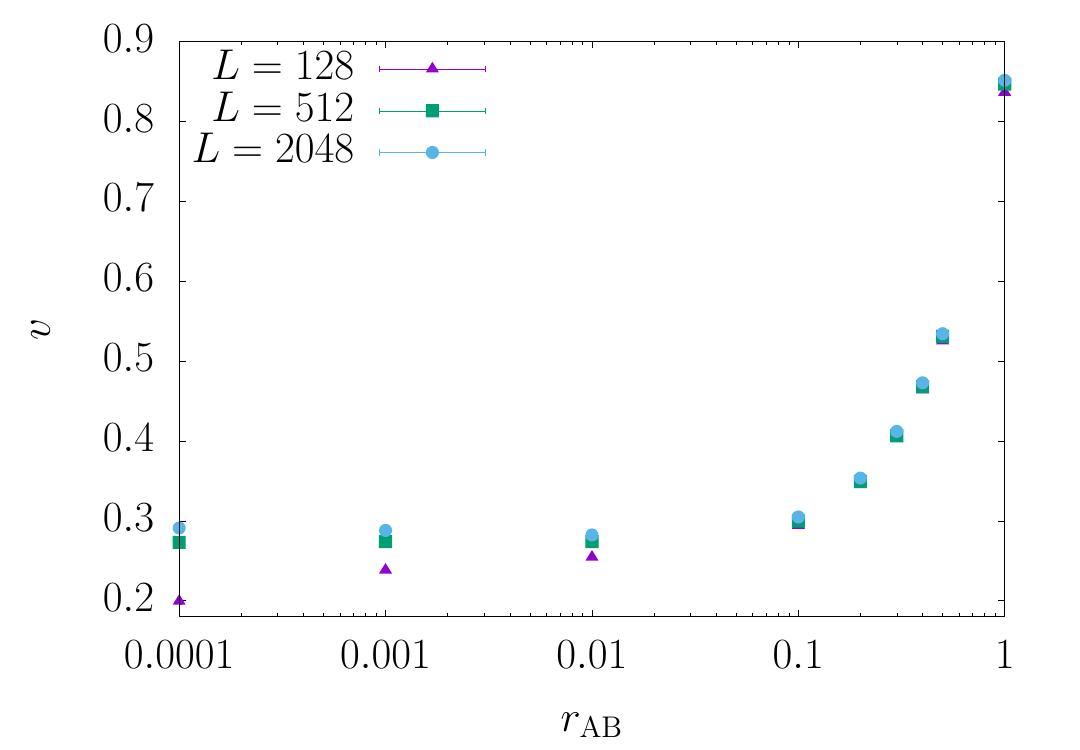}
    \caption{Front velocity as a function of $r_\mathrm{AB}$ and for different lateral sizes, namely $L = 128$, 256, and 2048. The error bars are smaller than the symbol size in all cases. All units are arbitrary.}
    \label{fig:vel}
\end{figure}

For all $r_\mathrm{AB}$, the average front grows linearly over time as 
\begin{equation}
    \langle \overline{h (t)} \rangle =v t+a_0,
    \label{eq:linear_front}
\end{equation}
where $v$ is the front velocity and $a_0$ is a parameter which depends on $r_\mathrm{AB}$. Figure \ref{fig:vel} plots the front velocity as a function of $r_\mathrm{AB}$ for systems with different lateral sizes $L$. Two distinct regimes are apparent in this plot. For $0~<~r_\mathrm{AB} \lesssim 0.1$, the front velocity remains approximately $r_\mathrm{AB}$-independent, whereas it increases with this parameter for $r_\mathrm{AB} \gtrsim 0.1$. At any rate, $v\neq 0$ for all these parameter values, hence the front is always in a moving phase. Incidentally, note that the front velocity does not depend significantly on $L$, except for very low $r_\mathrm{AB}$, for which $v$ increases slightly with $L$. 

\begin{figure}[htb]
    \centering
    \includegraphics[width=0.45\textwidth]{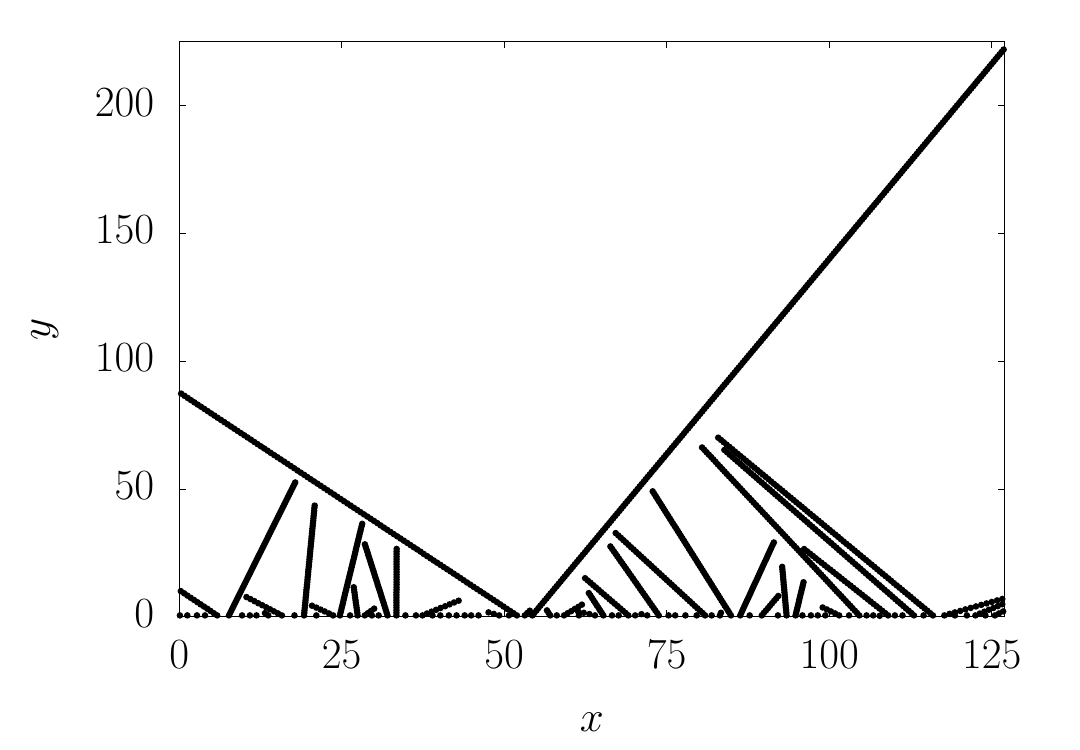}
    \caption{System with $L=128$ and $r_\mathrm{AB}=0$  for $t = 700$. Two branches have reached the boundary so that no further front growth is possible and the system freezes into the steady state. All units are arbitrary.}
    \label{fig:morfp0}
\end{figure}

\begin{figure}[htb]
    \centering
    \includegraphics[width=0.45\textwidth]{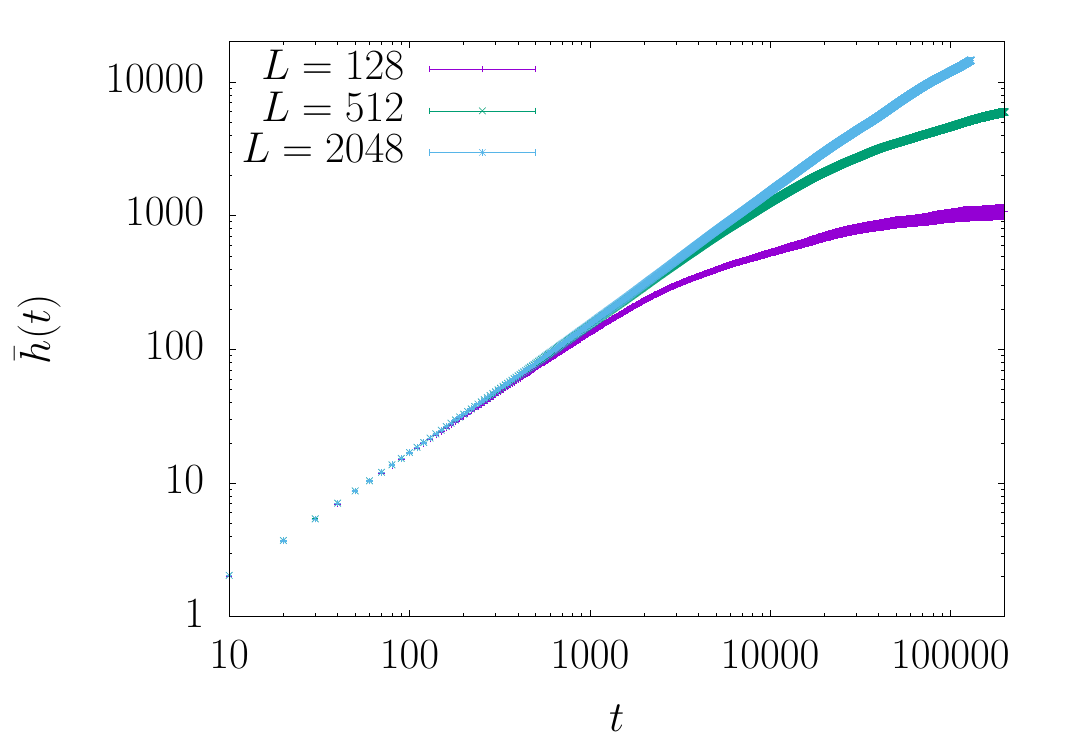}
    \caption{Mean front $\langle \overline{h(t)} \rangle$ as a function of time for $r_\mathrm{AB}=0$. The slopes of the curves are decreasing to zero as time increases. All units are arbitrary.}
    \label{fig:velp0}
\end{figure}

The $r_\mathrm{AB}=0$ case deserves a more detailed analysis \cite{Orozco2019}. By definition, only AA bonds form in this case. This means that secondary branches do not form and branches grow only along relatively few directions. As a result, just a few branches dominate the remaining ones and eventually reach the system edge, arresting the average growth of the front; Figure \ref{fig:morfp0} shows an example of such a situation. Therefore, for the $r_\mathrm{AB} =0$ case the system becomes pinned at steady state, being characterized by a zero average velocity, $v = 0$, for long enough times. This fact has been confirmed by computing the average front $\langle \overline{ h(t)} \rangle$ for $r_\mathrm{AB} = 0$ as a function of time, shown in Fig.~\ref{fig:velp0}. For $L = 128$, the front grows linearly with $t$ for short times, consistent with Eq.\ \eqref{eq:linear_front}, but the front velocity decreases monotonically for $t \gtrsim 1000$ to vanish asymptotically. The same trend is apparent for larger lateral sizes, but the linear growth regime extends to longer times as $L$ increases; in particular, the expected plateau for $L=2048$ is not observed because it requires computation times which are beyond our capabilities.

The increase of the velocity with the lattice size means (apart from pathological behaviors) that the velocity will reach a non zero asymptotic value for $r_\mathrm{AB}>0$. Hence, the picture that emerges from this analysis is a discontinuity of the front velocity $v$ suggesting a first order (discontinuous) transition: $v$ is different from zero if $r_\mathrm{AB}>0$ and equal to zero if $r_\mathrm{AB}=0$. The apparent change of the behavior of the velocity for $r_\mathrm{AB}\simeq 0.1$ (the flattening of its slope) can be interpreted as a crossover effect and not as the effect of a new phase transition.

Beyond aggregate morphologies and front velocity, the behavior of the front fluctuations, i.e., the kinetic roughening properties, turns out to also differ for $r_\mathrm{AB} = 0$ and for $r_\mathrm{AB} > 0$. Hence, we report the results obtained in both cases in the next sections. Unless otherwise stated, all calculations were carried out for $L = 2048$ and times $t\lesssim 20000-130000$. For comparison, in Ref.\ \cite{Dias2018} simulations were performed for system sizes $L\le 512$ and times $t\lesssim 4000$.

\subsection{Front roughness}

\begin{figure}[htb]
    \centering
    \includegraphics[width=0.45\textwidth]{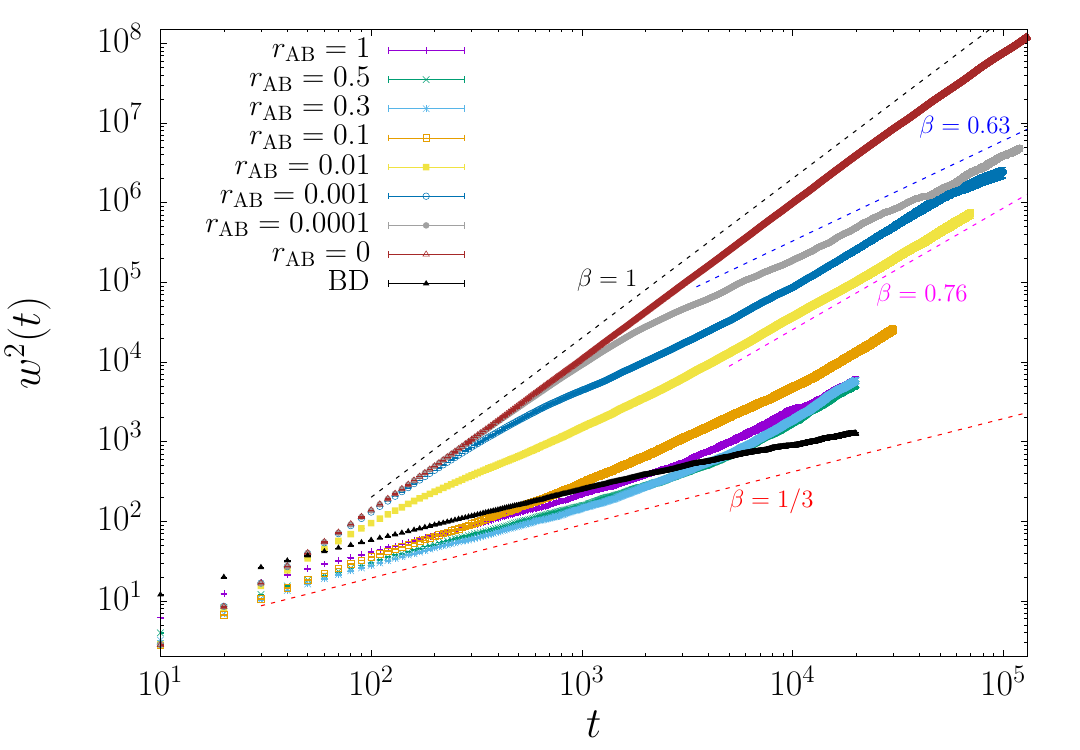}
    \caption{Squared front roughness vs time for values of $r_\mathrm{AB}$ as indicated in the legend. Error bars are smaller than the symbol size. All units are arbitrary.} 
    \label{fig:L2048curvasw2}
\end{figure}

Figure~\ref{fig:L2048curvasw2} plots the squared front roughness $w^2(t)$ calculated for several values of $r_\mathrm{AB}$. This figure shows that the front roughness increases with time in all cases. Moreover, note that the system sizes employed guarantee that the interface does not saturate into steady state for any value of $r_\mathrm{AB}$. Given the crossover behavior that is discussed below, this allows us to assess the most relevant mechanisms that control the large scale behavior of the system. For a qualitative analysis, curves in Fig.~\ref{fig:L2048curvasw2} have been fitted to the FV scaling law \eqref{eq:w} within two distinct time intervals, namely \textit{intermediate times} and \textit{long times}; very short times have been ignored. The results, reported in Table.~\ref{tab:beta}, exhibit two distinct regimes depending on time;  the rationale to consider intermediate times is to match with the results reported previously in the literature:
\begin{enumerate}
    \item  One observes a power-like trend for $t \lesssim 1000$ (intermediate times) and $r_\mathrm{AB} \ge 0.3$, which is consistent with the $\beta=1/3$ exponent for 1D-KPZ. On the other hand, for $r_\mathrm{AB}=0.01$ and $0.001$ the growth exponent value seems consistent with the $\beta\simeq 0.63$ QKPZ value. The $r_\mathrm{AB}=0.1$ case seems to be a crossover. See additional results in Table \ref{tab:beta}. Note that, for $r_\mathrm{AB}\le 0.01$, the $w^2$-curves overlap the $r_\mathrm{AB}=0$ curve at short times (see Fig.~\ref{fig:L2048curvasw2}). 
    
    \item For longer times, the curvatures increase regardless of $r_\mathrm{AB}$, suggesting very high asymptotic values for $\beta$. As an example, $\beta \simeq 0.76$ for $r_{\rm AB}=0.01$ and $t\gtrsim 3000$. See additional results in Table \ref{tab:beta}. As an additional reference, we have run simulations of the simple off-lattice ballistic deposition (BD) model, see Appendix \ref{app:BD}. The BD data are included in Fig.\ \ref{fig:L2048curvasw2} and feature the expected KPZ scaling for all simulated times. In particular, the difference with the $r_{\rm AB}=1$ case becomes apparent for $t\gtrsim 3000$.
    
    \item For $r_\mathrm{AB} = 0$, the value of $\beta \approx 1$ is essentially time-independent.
    
\end{enumerate}

\begin{table*}
\caption{Growth exponents $\beta$ for different values of $r_\mathrm{AB}$ computed for intermediate and long times.}
\label{tab:beta}
\begin{ruledtabular}
\begin{tabular}{dcdcd}
     r_\mathrm{AB} & intermediate times & \beta & long times & \beta \\\hline
     1 & $[40:800]$ & 0.349(7) & $[3000:20000]$ & 0.62(4)\\
     0.5 & $[100:1000]$ & 0.352(2) & $[5000:20000]$ & 0.76(4)\\
     0.3 & $[60:1000]$ & 0.353(5) & $[6000:20000]$ & 0.81(3)\\
     0.1 & $[100:1000]$ &  0.462(6) & $[15000:30000]$ & 0.80(5)\footnote{For $r_\mathrm{AB}=0.1$ there exists another time interval between those shown, namely $[1000:15000]$, with $\beta=0.62(2)$.}\\
     0.01 & $[350:2000]$ & 0.603(7) & $[5000:70000]$ & 0.76(3) \\
     0.001 & $[500:2000]$ & 0.594(10)  & $[2500:40000]$ & 0.73(2) \\
     0.0001 & $[60:900]$ & 0.920(2) & $[2000:120000]$ & 0.632(11) \\
     0 & $[200:30000]$ & 0.978(2) & $[70000:130000]$ & 0.87(6) \\
\end{tabular}
\end{ruledtabular}
\end{table*}

\subsection{Height-difference correlation function}
\label{sec:correlation_function}

\begin{figure*}[htb]
    \centering
    \includegraphics[width=1\textwidth]{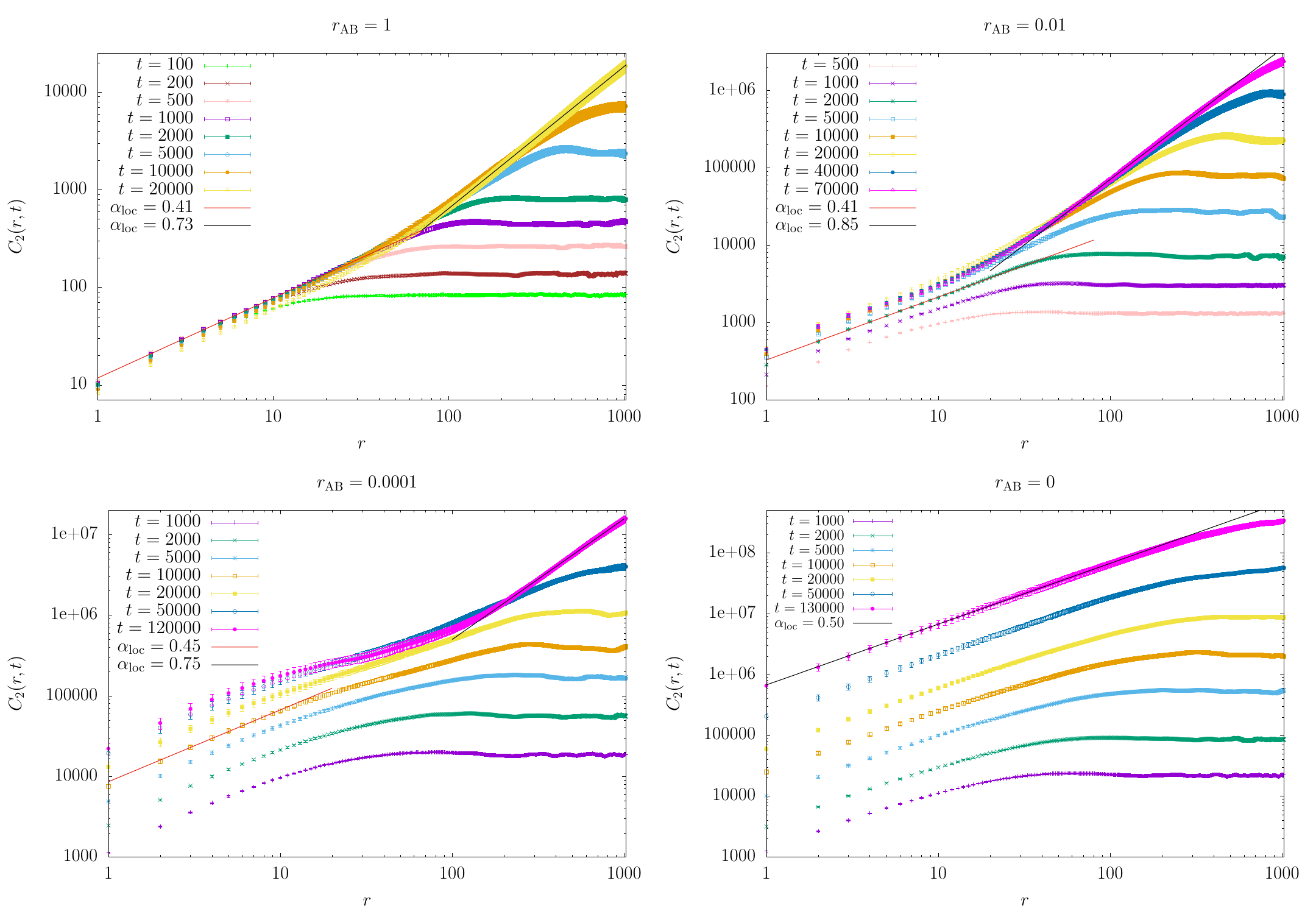}
    \caption{Front correlation function $C_2(r,t)$ vs $r$ for various $r_\mathrm{AB}$ and times. Solid lines correspond to fits to the power law $r^{2\alpha_{\mathrm{loc}}}$ within different $r$ intervals. All units are arbitrary.}
    \label{fig:c2}
\end{figure*}

We have computed the correlation function $C_2(r,t)$ for several times and $r_\mathrm{AB}$ values in the $[0,1]$ interval; the results are shown in Fig.~\ref{fig:c2}. The $C_2(r,t)$ function exhibits a relatively complex behavior, some of whose characteristics follow: 

\begin{itemize}
    \item As a general trend (noticeably, except for $r_\mathrm{AB} = 0$), the power law $C_2(r,t) \sim r^{2\alpha_{\rm loc}}$ holds for all $r_\mathrm{AB}$ at short distances relative to the correlation length $\xi$, while a plateau is reached at larger distances. However, for long enough times and $r_\mathrm{AB} \ne 0$, a second power law can be measured in the region of large $r$ featuring an $\alpha_{\mathrm{loc}}$ exponent larger than that found at short distances; this effect is highlighted by straight lines with different slopes in Fig.~\ref{fig:c2}. As examples, the $C_2(r,t)$ curve for $r_\mathrm{AB} = 1$ exhibits $\alpha_\mathrm{loc}$ = 0.41 for $r \lesssim 100$; the value of $\alpha_\mathrm{loc}$ increases for larger $r$ to reach 0.73 for the longest time reached. For $r_\mathrm{AB} = 0.01$, one gets $\alpha_\mathrm{loc}$ values as high as 0.85 at large distances. The existence of two different scaling behaviors for small and large distances at long times does not take place in off-lattice BD, for which $C_2(r,t)$ features a single roughness exponent $\alpha_{\rm loc} = \alpha=1/2$ for all values of $r$ and $t$; see Fig.~\ref{fig:BD_c2} in Appendix \ref{app:BD}.

    \item For very small $r_\mathrm{AB}$, as $0.0001$ (see Fig.~\ref{fig:c2} bottom left), the $C_2(r,t)$ curves show a different behavior in between short and large distances for very long times. More details about the potential origin of this unusual behavior will be given in Sec.~\ref{forma_limite} below.

    \item For each $0 \le r_\mathrm{AB} \le 0.01$, the $C_2(r,t)$ curves shift steadily upwards and do not overlap for increasing times, until the second power-law regime eventually appears; from then on, the curves approximately overlap for small $r$. This behavior is indicative of anomalous kinetic roughening \cite{Lopez1997,Ramasco2000,Cuerno2004}. For $r_\mathrm{AB} \ge 0.1$, such a vertical shift of the curves over time is not apparent (see Fig.~\ref{fig:c2}, $r_\mathrm{AB} = 1$ case). This could be understood accepting that, for short times and large $r_\mathrm{AB}$, the system is in the KPZ universality class, which does not exhibit anomalous scaling. A more detailed discussion about the anomalous scaling will be reported in Sec.~\ref{anomalous_scaling} below. 
\end{itemize}

\subsection{Correlation length}
In Sec.~\ref{sect:observables} we described the method to calculate the correlation length $\xi(t)$ from the height-difference correlation function. This approach can be applied only if each $C_2(r,t)$ curve grows with $r$ until the correlation length is surpassed and the correlation function becomes constant. As $\xi(t)$ increases with time, the method can no longer be used when it becomes comparable to the system size. Moreover, as we already mentioned, at long times the $C_2(r,t)$ curves do not exhibit standard behavior in this system. Consequently, the method described in Sec.~\ref{sect:observables} to compute $\xi(t)$ is applicable only within the range of intermediate times. 

\begin{figure}[htb]
    \centering
    \includegraphics[width=0.45\textwidth]{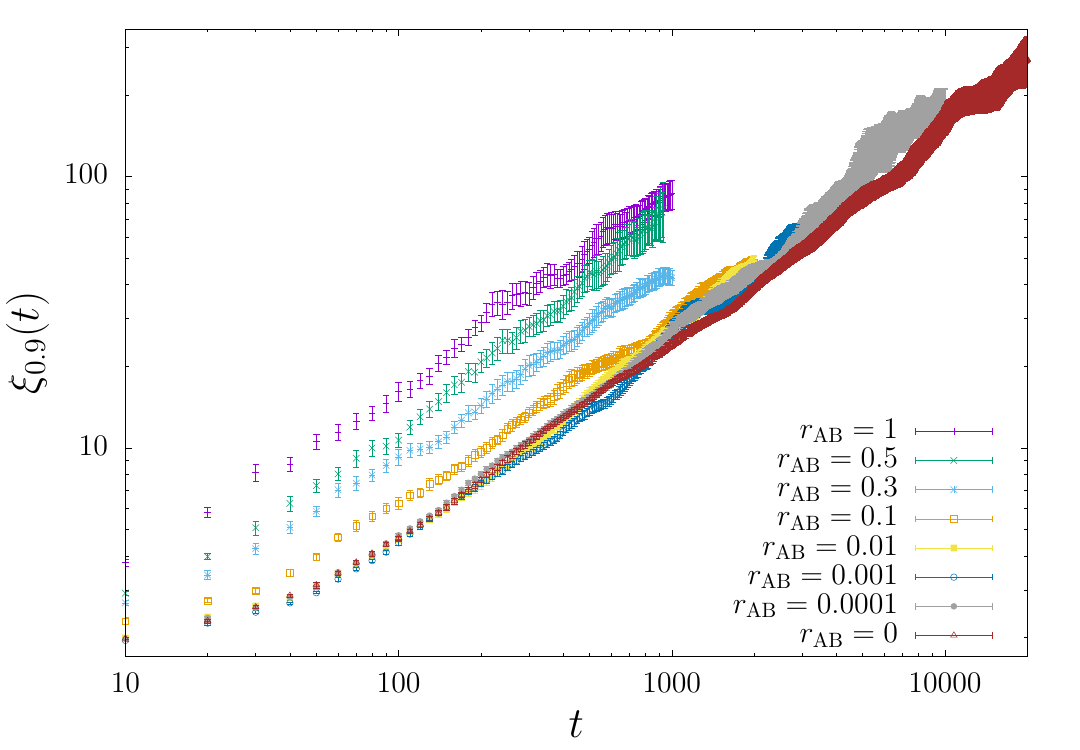}
    \caption{Correlation length vs time for several values of $r_\mathrm{AB}$, as indicated in the legend. All units are arbitrary.}
    \label{fig:L2048z}
\end{figure}

Fig.~\ref{fig:L2048z} shows the correlation length computed from Eq.~\eqref{eq:compute_corr_length} with $a=0.9$, $\xi_{0.9}(t)$, as a function of time for different values of $r_\mathrm{AB}$. The data may be rationalized by fitting the $\xi(t)$ curves to Eq.~\eqref{eq:correlation_length}, which allows one to compute the dynamic exponent $z$ for each $r_\mathrm{AB}$. Our quantitative results, displayed in Table~\ref{tab:exponents_z_alpha}, are compatible within error estimates with the KPZ universality class ($z_\mathrm{KPZ}=3/2$) for $r_\mathrm{AB}\geq 0.1$, and with the shared $z\approx 1$ value of the QKPZ and moving DPD phase universality classes, for $0 < r_\mathrm{AB}\le 0.01$. To our knowledge no previous reports are available on the value of the $z$ exponent for the model of patchy colloids.

\begin{figure}[htb]
    \centering
    \includegraphics[width=0.45\textwidth]{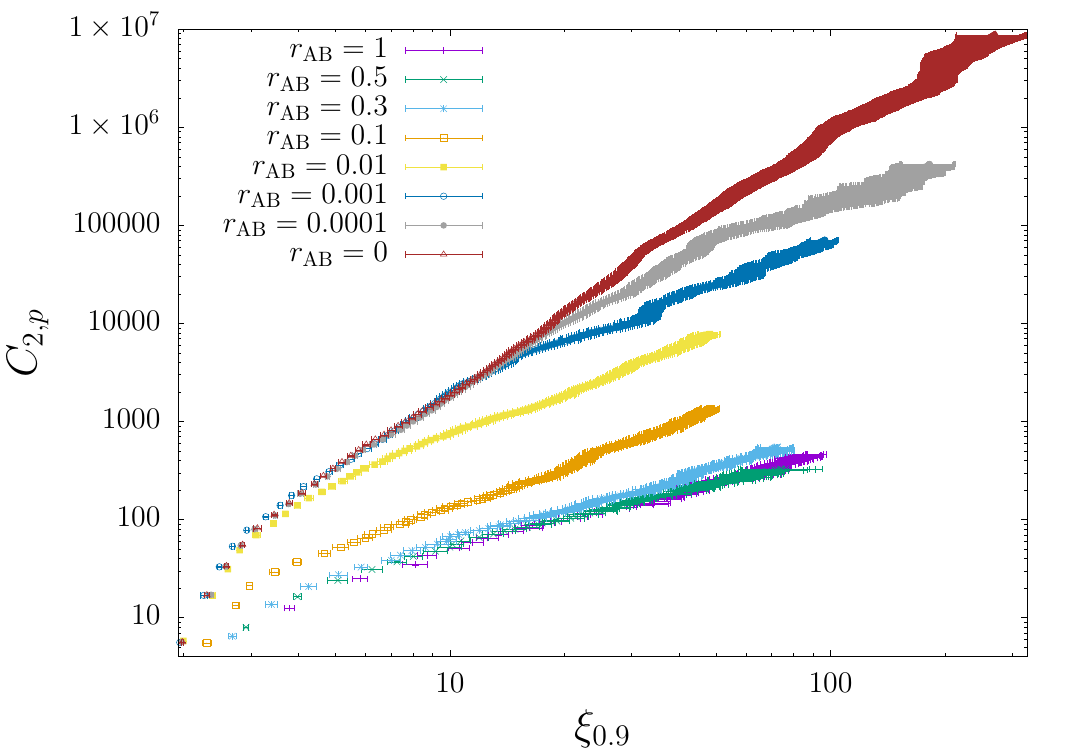}
    \caption{Plateau correlation value vs correlation length for values of $r_\mathrm{AB}$ as indicated in the legend. All units are arbitrary.}
    \label{fig:L2048alpha}
\end{figure}

The global roughness exponent $\alpha$ can also be computed by plotting the plateau of the height-difference correlation function, $C_{2,p}(t)$, vs the correlation length $\xi_{0.9}(t)$ \cite{Barreales2020}, as shown in Fig.~\ref{fig:L2048alpha}. The $\alpha$ values calculated by fitting such a plot to Eq.~\eqref{eq:corr_length} are shown in Table~\ref{tab:exponents_z_alpha} as well. For $r_\mathrm{AB} \ge 0.3$  the estimates of the roughness exponent are again in agreement with the 1D-KPZ universality class ($\alpha_\mathrm{KPZ}=1/2$). For $0 < r_\mathrm{AB} \le 0.1$, the roughness exponent value is compatible with the 1D-DPD universality class in the moving phase. For $r_\mathrm{AB}=0$, both the roughness and the dynamic exponents deviate from the previous values and are larger than one. 

Since $\beta = \alpha/z$, one may compare the ratio $\alpha/z$ from Table~\ref{tab:exponents_z_alpha} with the $\beta$ values from Table~\ref{tab:beta}, computed directly from the front roughness. The ratios $\alpha/z$ are compatible with the growth exponent measured at intermediate times, except for $r_\mathrm{AB} =0.001$ and $r_\mathrm{AB} =0.0001$, for which which $\alpha/z$ values are closer to the $\beta$ obtained at long times. This suggests that, for $r_\mathrm{AB} > 0.001$, the asymptotic regime has not been reached yet, and the growth regimes shown in Fig.~\ref{fig:L2048curvasw2} are still intermediate times.

\begin{table*}[htb]
\caption{Critical exponents $z$ and $\alpha$ for several values of $r_\mathrm{AB}$.}
\begin{ruledtabular}
\begin{tabular}{dcddd}
     r_\mathrm{AB} & intermediate times 
     & z & \alpha &\alpha/z \\\hline
     1 & [10:1000] & 1.44(6) & 0.50(2) & 0.35(2)\\
     0.5 & [100:1000] & 1.20(11) & 0.45(3) & 0.37(4)\\
     0.3 & [20:1000] & 1.46(5) & 0.54(2) & 0.37(2)\\ 
     0.1 & [30:2000] & 1.47(4) & 0.721(16) & 0.489(18)\\
     0.01 & [200:1500] & 1.22(5) & 0.73(3) & 0.60(3) \\ 
     0.001 & [2000:5000] & 1.11(15) & 0.77(7) & 0.69(11) \\ 
     0.0001 & [3000:10000] & 1.07(22) & 0.71(14) & 0.66(19) \\
     0 & [1400:20000] & 1.18(8) & 1.18(7) & 1.00(9)\\ 
\end{tabular}
\end{ruledtabular}
\label{tab:exponents_z_alpha}
\end{table*}

\subsection{Intrinsic anomalous scaling} 
\label{anomalous_scaling}

The fact that the height-difference correlation function $C_2(r,t)$ curves shift steadily upwards with increasing time is an indication of anomalous kinetic roughening behavior \cite{Lopez1997,Ramasco2000,Cuerno2004}. As we mentioned in Sec.~\ref{sec:correlation_function}, this behavior is observed for $r_{\mathrm{AB}}\le 0.01$.  

\begin{figure}[htb]
    \centering
    \includegraphics[width=0.45\textwidth]{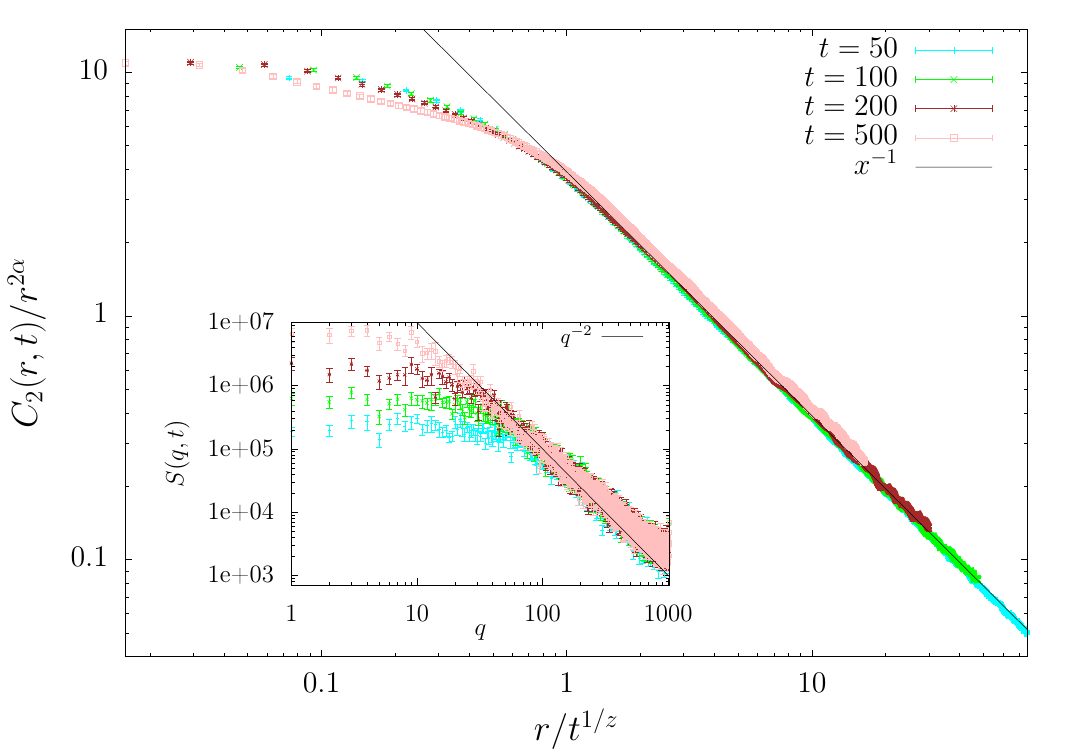}
    \caption{Data collapse for intermediate times of $C_2(r,t)$ using KPZ exponents $z=1.5$ and $\alpha=0.5$ for $r_{\rm AB}=1$. Solid line correspond to the power law  $g(x)\sim x^{-2\alpha}$ for $x>1$ with  $x = r/t^{1/z}$, Eq.~\eqref{eq:corr_length}.  Inset: Structure factor for the same data. Solid line corresponds to $q^{-(2\alpha+1)}$. All units are arbitrary.}
    \label{fig:c2_scaling_p1}
\end{figure}

In a system ruled by the FV scaling Ansatz, the scaling function $g_{\rm FV}(u)$ appearing in Eq.~\eqref{eq:corr_lengthFV} does not depend on $u$ for $u\ll 1$, e.g.\ the KPZ equation \cite{Barabasi1995,Krug1989}. As discussed in the previous sections, we obtain KPZ scaling exponents at intermediate times for large $r_{\mathrm{AB}}$, and we have inquired what type of scaling Ansatz ---whether FV, as expected for the KPZ equation, or an anomalous one--- occurs in this regime. An example of this type of analysis is shown in Fig.~\ref{fig:c2_scaling_p1}, which displays the intermediate-time behavior of the scaled front height-difference correlation function ($C_2(r,t)/r^{2\alpha}$ vs $r/t^{1/z}$) and structure factor ($S(q,t)$ vs $q$, inset) for $r_{\rm AB}=1$. This behavior is largely consistent with a FV Ansatz, i.e., scaling is not anomalous for intermediate times. 

\begin{figure}[htb]
    \centering
    \includegraphics[width=0.45\textwidth]{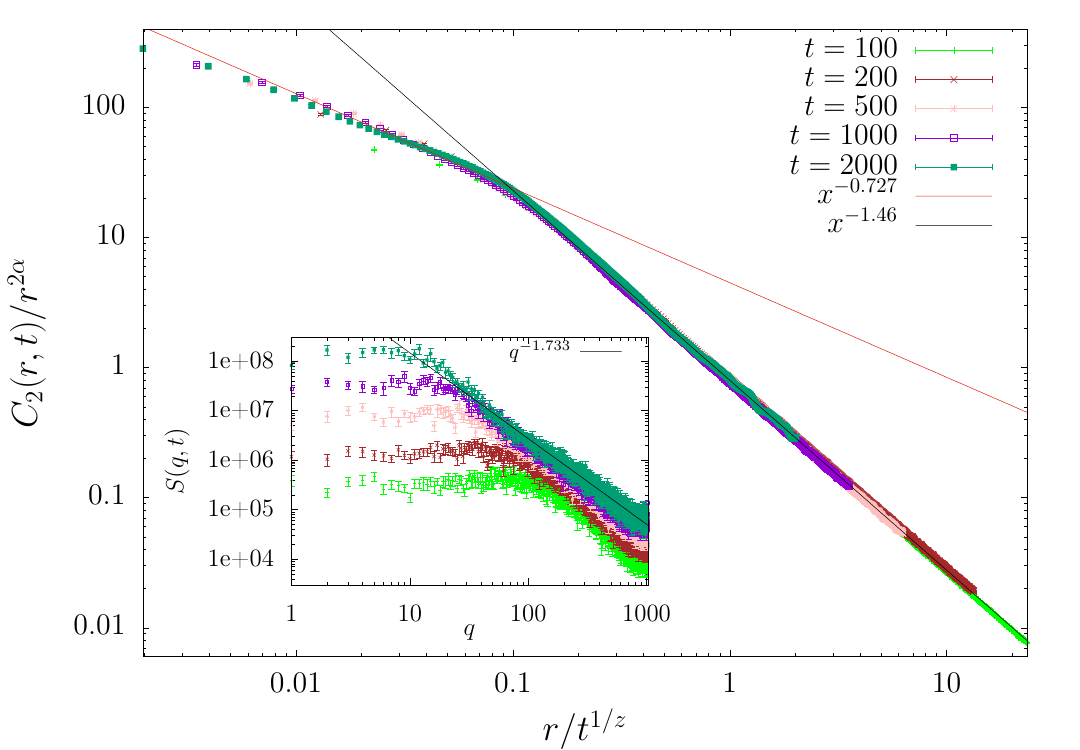}
    \caption{Data collapse of $C_2(r,t)$ using the computed exponents $z=1.22$ and $\alpha=0.73$ for $r_{\rm AB}=0.01$. Solid lines correspond to the power laws $g(x)\sim x^{-2\alpha'}$ (fitting to the $t=1000$ curve for $x<0.1$) and $g(x)\sim x^{-2\alpha}$,  with  $x = r/t^{1/z}$.  Inset: Structure factor for the same data. Solid line corresponds to $q^{-(2\alpha_{\mathrm{loc}}+1)}$ where $\alpha_{\mathrm{loc}}=\alpha-\alpha'=0.366(4)$ as obtained from the analysis of the data collapse for $C_2(r,t)$. All units are arbitrary.}
    \label{fig:c2_scaling_p001}
\end{figure}

On the contrary, data computed for $r_\mathrm{AB} \le  0.01$ fit to a scaling function $g(u) \sim u^{-2(\alpha - \alpha_{\rm loc})}$, as in Eq.\ \eqref{eq:corr_length} with $\alpha_{\rm loc} \neq \alpha$, which, combined with $\alpha_{\rm loc} <1$, implies intrinsic anomalous scaling~\cite{Lopez1997}. As a representative example, Fig.~\ref{fig:c2_scaling_p001} shows a data collapse consistent with Eq.\ \eqref{eq:corr_length}, obtained using $C_2(r,t)$ data for $t=100, 200, 500, 1000, 2000$, and $r_\mathrm{AB} = 0.01$; within the time range covered by the plot, our data are compatible with $z=1.22(5)$ and $\alpha=0.73(3)$, as discussed above. Further, by fitting the $t=1000$ curve to $u^{-2\alpha'}$ for $u\ll 1$, with $\alpha' = \alpha - \alpha_{\rm loc}$, we obtain $\alpha'=0.364(4)$ and  therefore $\alpha_{\rm loc}=0.366(4)\neq \alpha$. 

The occurrence of intrinsic anomalous scaling can be further verified from the behavior of the front structure factor \cite{Lopez1997}, presented in the inset of Fig.~\ref{fig:c2_scaling_p001} for $r_\mathrm{AB} = 0.01$. The $S(q,t)$ curves show again a systematic shift upwards with increasing time, unambiguously attributable to intrinsic anomalous scaling~ \cite{Lopez1997}. As mentioned in Sec.~\ref{sect:observables} above, in such a case the structure factor scales with the wave vector as $S(q,t) \sim q^{-(2\alpha_{\rm loc}+1)}$ at large $q$, which depends on the local roughness exponent, see Eq.~\eqref{eq:scaling_s}. The straight line in the inset of Fig.~\ref{fig:c2_scaling_p001} corresponds to the $\alpha_{\rm loc}$ value obtained from the collapse of $C_2(r,t)$, as expected, so that our interpretation of the data in terms of intrinsic anomalous scaling is consistent. 

\begin{figure}[htb]
    \centering
    \includegraphics[width=0.45\textwidth]{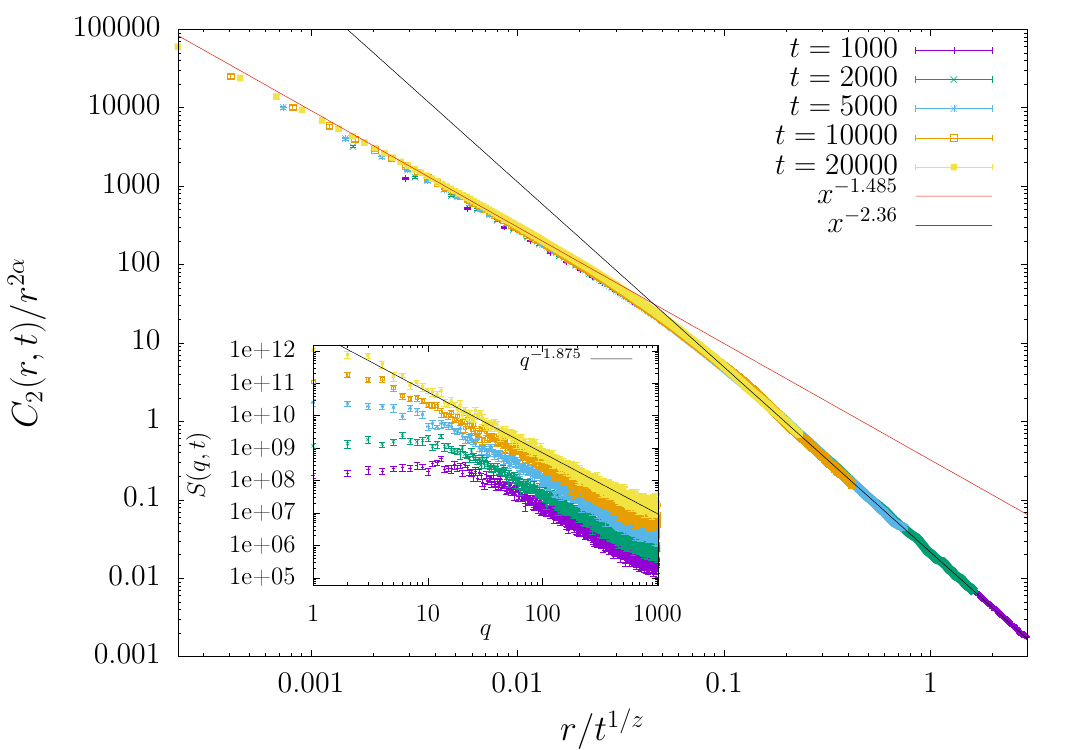}
     \caption{Data collapse of $C_2(r,t)$ using the computed exponents $z=1.18$ and $\alpha=1.18$ for $r_{\rm AB}=0$. Solid lines correspond to the power laws $g(x)\sim x^{-2\alpha'}$  (fitting to the $t=20000$ data for $x<0.03$) and $g(x)\sim x^{-2\alpha}$, with  $x=r/t^{1/z}$.  Inset: Structure factor for the same data. The solid line corresponds to $q^{-(2\alpha_{\mathrm{loc}}+1)}$ where $\alpha_{\mathrm{loc}}=\alpha-\alpha'=0.437(9)$, as obtained from the analysis of the data collapse for $C_2(r,t)$. All units are arbitrary.}
    \label{fig:c2_scaling_p0}
\end{figure}

We have also observed intrinsic anomalous scaling for $r_{\rm AB}=0$, which we consider next as a particularly interesting case. Indeed, visual inspection of the uncollapsed  $C_2(r,t)$ data in the fourth panel of Fig.~\ref{fig:c2} is suggestive of non-negligible anomalous scaling. The set of computed exponents [$z=1.18(8)$ and $\alpha=1.18(7)$] indeed yields a consistent data collapse, as specifically shown in Fig.~\ref{fig:c2_scaling_p0}. By fitting the scaling function for $t=20000$ we obtain $\alpha'=0.743(9)$, resulting into $\alpha_{\mathrm{loc}}=0.437(9)\neq \alpha$. As a final consistency check, the inset of Fig.~\ref{fig:c2_scaling_p0} shows the front structure factor $S(q,t)$ for $r_{\rm AB}=0$. The straight line in this inset corresponds to the  $q^{-(2\alpha_\mathrm{loc}+1)}$ behavior expected in this case, where the value of $\alpha_{\mathrm{loc}}$ computed from the height-difference correlation function has been used.

\subsection{Macroscopic shapes for $r_{\rm AB} \neq 0$}
\label{forma_limite}

Previously, we noted that the height-difference correlation function behavior for $r_{\rm AB} \neq 0$ becomes more complex at very long times. As shown in Fig.~\ref{fig:c2}, we can identify two different slopes, each for small and large $r$, in the log-log plot of $C_2(r,t)$ vs $r$. Hence, one might calculate two local roughness exponents $\alpha_{\mathrm{loc}}$ for the corresponding ranges of distances. Furthermore, the exponent $\beta$ measured from the front roughness increases its value at long times (see Table~\ref{tab:beta}). 

\begin{figure}[htb]
    \centering
    \includegraphics[width=0.45\textwidth]{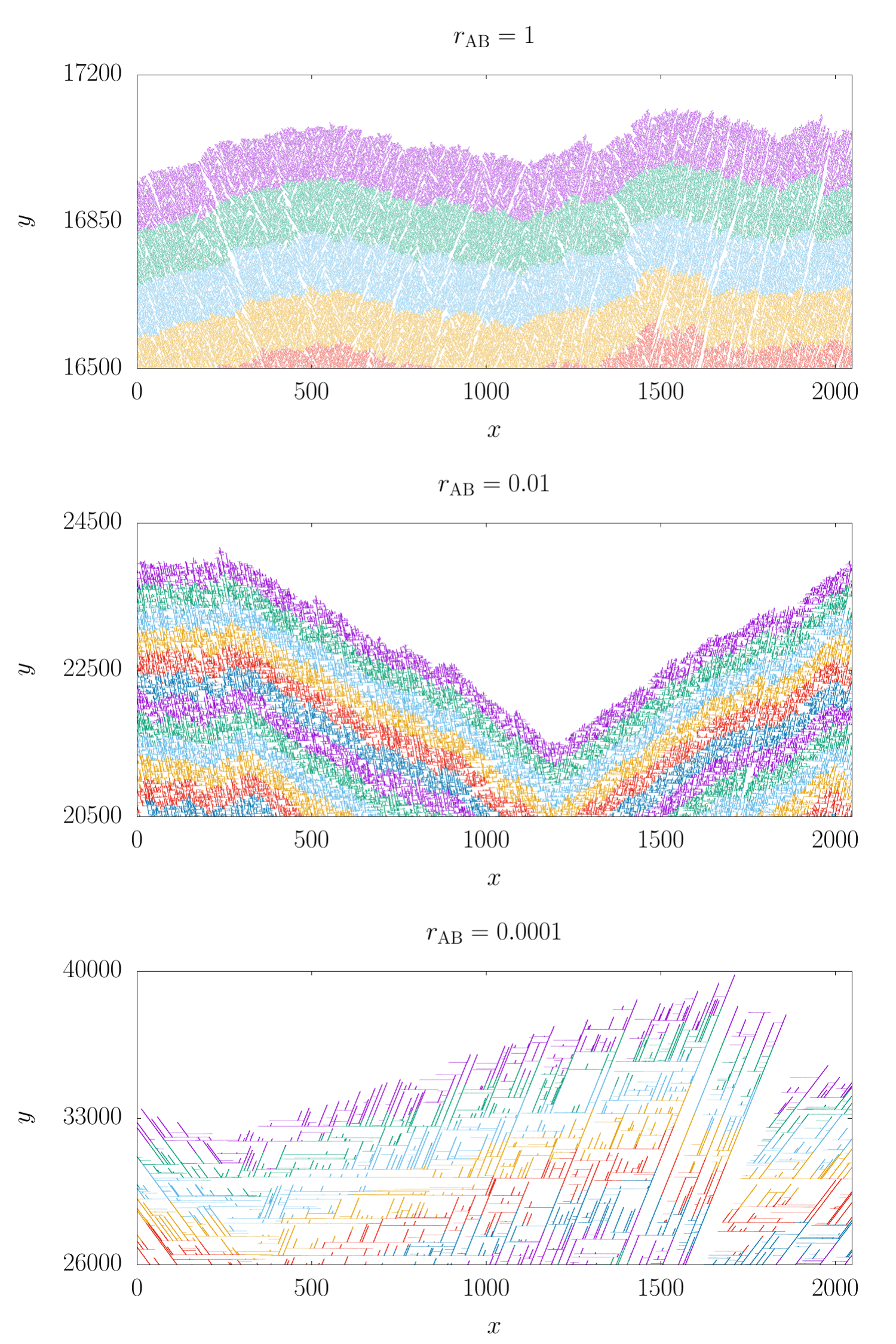}
    \caption{Morphologies of the colloidal aggregates at the final time, namely $t=20000$, $70000$, and $120000$, for $r_{\rm AB}=1$, $0.01$, and $0.0001$, respectively. Each color shows the last hundred thousand particles to join the system. Colors are arbitrary and their sole purpose is to show the different groups of $10^5$ particles, to see the shape of the aggregate front. All units are arbitrary.}
    \label{fig:morf_final}
\end{figure}

An explanation for these changes can be found in the time evolution of the colloidal aggregates. At early times, the fronts are morphologically isotropic for all $r_{\mathrm{AB}}$, as can be observed in Fig.~\ref{fig:morf_prob}, and the aggregate fronts fluctuate at distances much smaller than the system size. However, for longer times the aggregate ``splits'' into a few components of lateral sizes comparable to $L$, and the front displays macroscopic shapes. Some representative examples are shown in Fig.~\ref{fig:morf_final}. Note e.g.\ the development of large ``facets'' for $r_{\rm AB}=0.01$.

\begin{figure}[htb]
    \centering
    \includegraphics[width=0.45\textwidth]{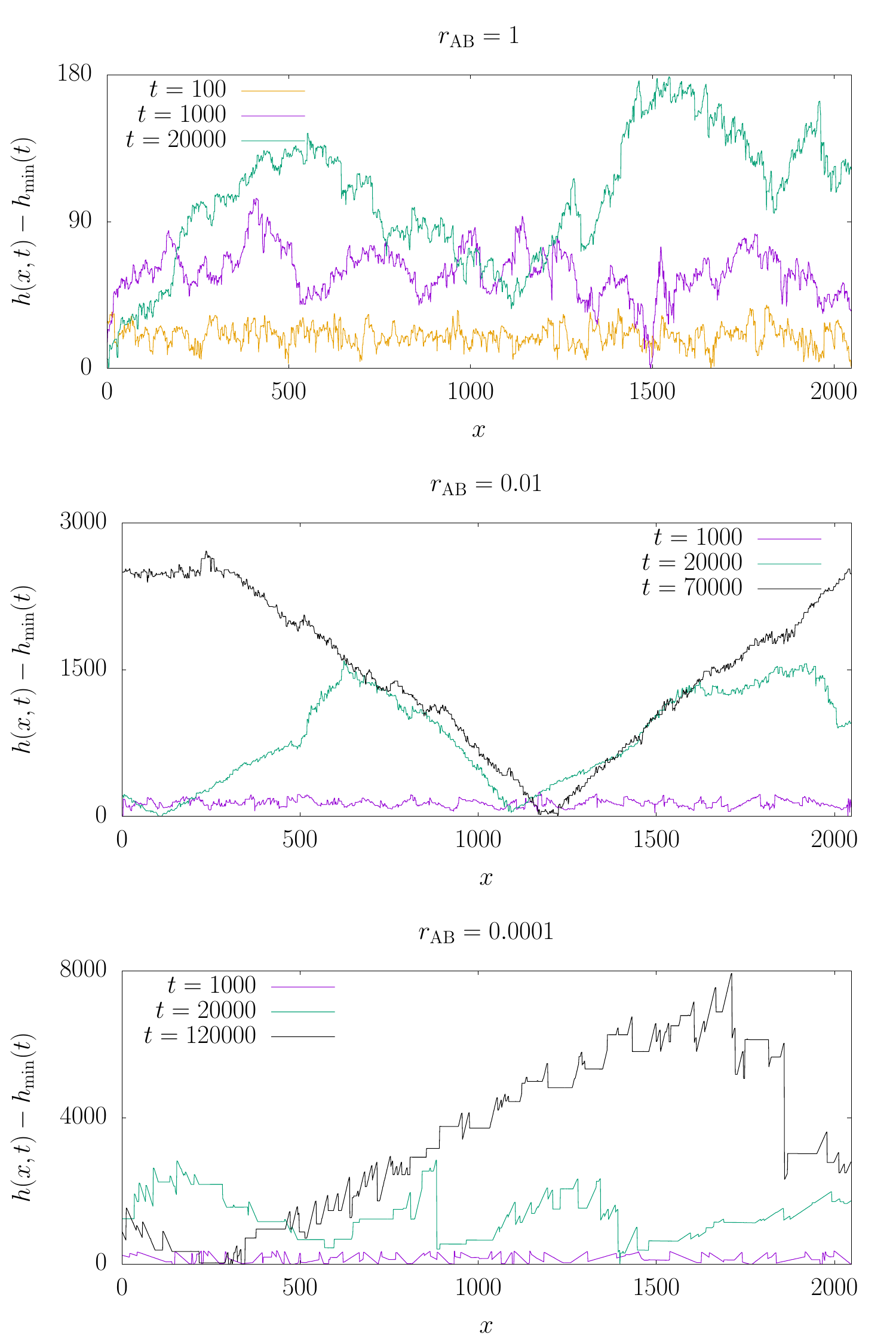}
    \caption{Temporal evolution of the height front $h(x,t)$ for $r_{\rm AB}=1$, $0.01$, and $0.0001$, top to bottom. The aggregates are the same as those shown in Fig.~\ref{fig:morf_final}. 
    All units are arbitrary.}
    \label{fig:evolucion_frente}
\end{figure}

We have observed that remarkable jumps of the front heights appear at those times at which the aforementioned second regime of the $C_2(r,t)$ curves appears. This effect is shown in  
Fig.~\ref{fig:evolucion_frente}, where we have represented the time evolution of $h(x,t)$ for the same conditions (i.e., $r_\mathrm{AB}$ values) as in Fig.\ \ref{fig:morf_final}. From this figure, one observes that the front displays small fluctuations at short times ($t \lesssim 1000$), none of which stands out from the average. At longer times, the mentioned jumps appear, resulting again into macroscopic shapes. This effect becomes stronger as $r_\mathrm{AB}$ decreases, and becomes evident only for sufficiently large values of $L$.

\subsection{Front structure for $r_{\rm AB}=0$}
\label{forma_limite_r0}

We have already pointed out that our model's behavior changes quite substantially when $r_{\rm AB}=0$. Recall also that the front velocity becomes zero for sufficiently long times for this parameter condition. Nevertheless, prior to macroscopic pinning, the kinetic roughening of the front turns out to share many features with that of an unstable generalization of on-lattice ballistic deposition \cite{Asikainen2002b}.

\begin{figure}[t]
    \includegraphics[width=0.45\textwidth]{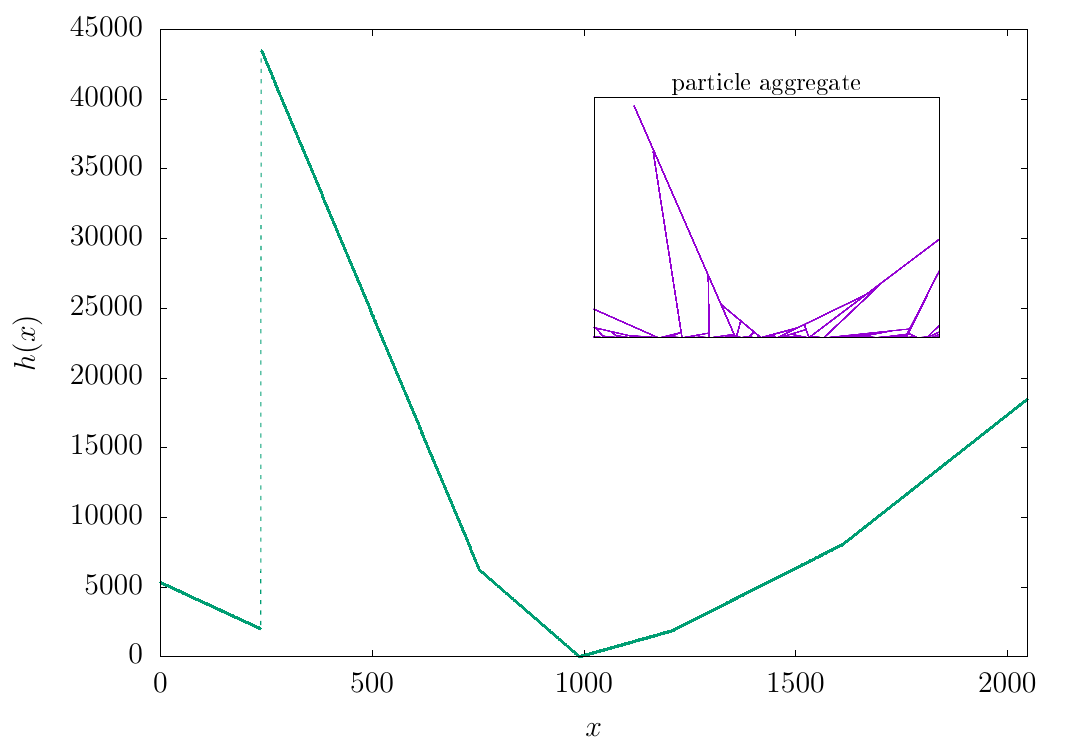}
    \caption{Sample front morphology obtained at $t=130000$ for $r_{\rm AB}=0$. The main panel shows the front height $h(x,t)$ as points connected by a dashed line. Note the very different scales employed for the two axes. The corresponding particle aggregate is shown in the inset. All units are arbitrary.}
    \label{fig:morf_r0}
\end{figure}

Indeed, for $r_\mathrm{AB}=0$ the front velocity eventually becomes zero due to the preferential growth of a few branches which span the full lateral system size, a behavior which is unlike that found at long times for $r_{\rm{AB}}>0$. Nevertheless, the peculiar form of the particle aggregate for $r_\mathrm{AB} = 0$ leads to a front which features very large values of the front derivative, or slope, at many of its locations, similar to what was seen for very small but nonzero $r_{\rm AB}$ values. An example of this is shown in Fig.~\ref{fig:morf_r0}.

\begin{figure}[t]
    \includegraphics[width=0.45\textwidth]{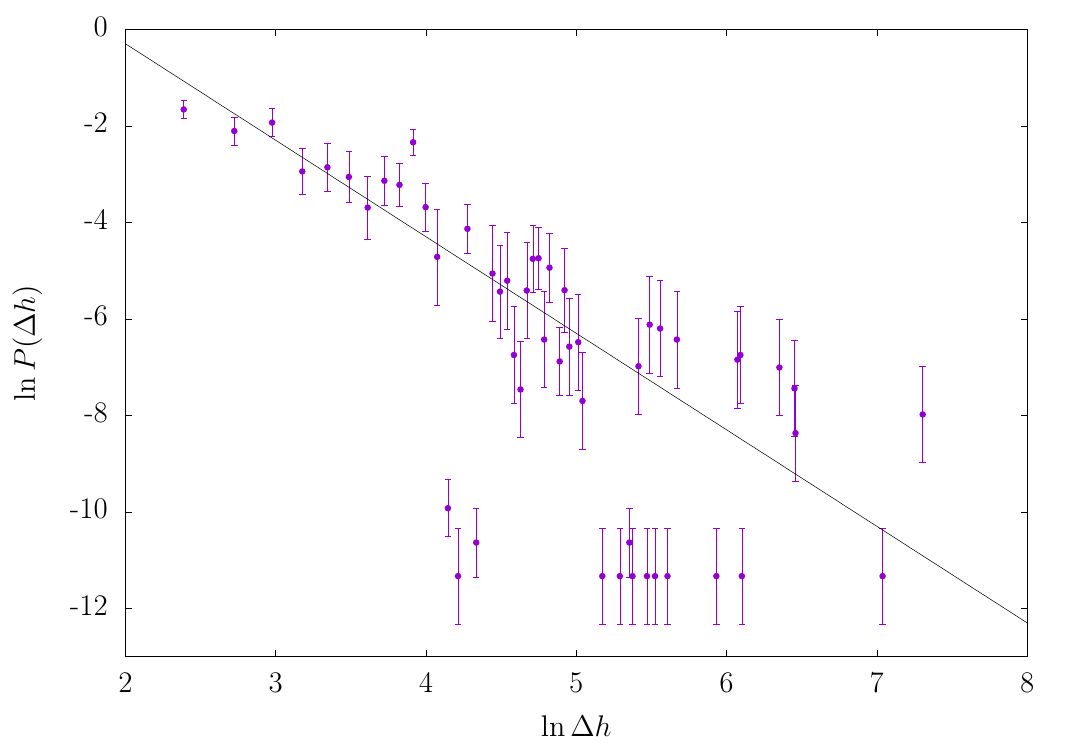}
    \caption{Slope histogram for the fronts obtained at $t=130000$ for $r_{\rm AB}=0$. For comparison, the solid line corresponds to $P(\Delta h) \propto 1/(\Delta h)^2$. All units are arbitrary.}
    \label{fig:histo_r0}
\end{figure}

Quantitatively, a very similar behavior to our $r_{\rm AB}=0$ case has been reported for a growth model that generalizes on-lattice ballistic deposition \cite{Asikainen2002b}. Specifically, while in standard BD 
particles are deposited vertically at random positions, in Ref.\ [\onlinecite{Asikainen2002b}] particles follow straight trajectories with a random inclination with respect to the $x$ axis, whose angle is chosen from an uniform distribution. Such a modification of BD is known to lead to a morphological instability
\cite{Bales1990,Meakin1998} inducing the formation of large columns. Indeed, in the simulation of Ref.\ [\onlinecite{Asikainen2002b}]
the front was found to feature large values of the slope $\Delta h$ at many places, to the extent that the probability distribution function $P(\Delta h)$ for the slope values decreased slowly as a power law of the form $P(\Delta h) \sim 1/(\Delta h)^{\gamma}$,
with a value for the $\gamma$ exponent between 1.6 and 2; for very large $\Delta h$, the distribution falls off much faster. 
This behavior was argued to lead to intrinsic anomalous scaling \cite{Asikainen2002,Asikainen2002b}. Remarkably, the scaling exponents measured in Ref.\ [\onlinecite{Asikainen2002b}] for this unstable BD model, namely, $\alpha\simeq 1.25$, $\alpha_{\rm loc}\simeq 0.54$, $\beta\simeq 1$, and $z\simeq 1.25$, are quite close to those we presently obtain for the patchy colloid model for $r_{\rm AB}=0$ prior to pinning, namely, $\alpha \simeq 1.18$, $\alpha_{\rm loc} \simeq 0.45$, $\beta\simeq 0.98$, and $z\simeq 1.18$. In view of this, we have estimated the slope of the $P(\Delta h)$ histogram obtained herein for $r_{\rm AB}=0$; the results, shown in Fig.~\ref{fig:histo_r0} are not so far from those reported in Ref.\ [\onlinecite{Asikainen2002b}]. The arguments employed in this reference can be similarly employed here to explain the intrinsic anomalous scaling for $r_{\rm AB}=0$. Quantitative differences with the unstable BD model \cite{Asikainen2002b} are probably induced by the deviations of our slope histogram from the $P(\Delta h)$ function obtained for the latter, especially at the largest $\Delta h$ values. As a conclusion, we believe that this result may be indicative of a morphological instability analogous to that of oblique-incidence BD, to be responsible for the intrinsic anomalous scaling of the fronts of the colloidal aggregates formed for $r_{\rm AB}=0$.

\section{Discussion}
\label{sec:disc}

In this paper, we have revisited the model of ``patchy'' colloids for the coffee-ring effect developed by Dias {\em et al}.\ \cite{Dias2013,Dias2014,Dias2018}. For such a purpose, we have extended the numerical simulations to longer times in larger systems. We have also used more realistic boundary conditions and an extended characterization in terms of correlation functions computed in the real and reciprocal spaces. This results into a wider overlook of the problem, from which new conclusions may be stated.

Let us consider first the set of scaling exponents for each condition, i.e., $r_\mathrm{AB}$ value. In general, the results reported herein are in fair agreement with those by Dias {\em et al}.\ for comparable simulation times and sizes. Deviations appear for longer times, though. In particular, the $\alpha$ and $\beta$ exponent values which were ascribed earlier to the QKPZ universality class are now seen to be crossover values limited to the intermediate-time evolution in a restricted range of $r_{\rm{AB}}$ values.

Taking into account our full set of results including very small $r_{\rm AB}\geq 0$ 
and long times, the overall behavior seems already clear in Fig.\ \ref{fig:L2048curvasw2}. We can classify the behavior in terms of large or small $r_{\rm AB}$, with the boundary being approximately at $r_{\rm AB}=0.01$. For each value of this parameter, we have to distinguish between intermediate and long times. Thus, for large $1 \geq r_{\rm AB} > 0.01$ and intermediate times, exponent values are KPZ for $r_{\rm AB}=1$ and gradually increase (going through QKPZ-like values) for decreasing $r_{\rm AB}$. However, at longer times $\beta$ and $\alpha$ both increase substantially, even for the $r_{\rm AB}=1$ case, which hence differs from simple off-lattice BD. 

At this, recall that, for models with time-dependent (rather than quenched) noise like for these patchy colloids, $\beta>1/2$ is usually taken as an indication of some morphological instability, since $\beta=1/2$ corresponds to purely random deposition of particles \cite{Barabasi1995,Meakin1998}. Crossover in time from e.g.\ KPZ-like scaling behavior into a regime featuring much larger, effective exponent values is known in other growth systems with time-dependent noise. An important example is that of diffusion-limited growth systems, see e.g.\ Ref.\  \cite{Nicoli2009} and other therein. 

For instance, the fronts of bacterial colonies growing in a parameter region expected to correspond to an Eden-like (i.e., KPZ-like) \cite{Meakin1998} behavior have been recently seen in experiments and models \cite{Santalla2018} to change with time from a compact regime with relatively small fluctuations into a long time regime dominated by a branched morphology with very large slopes. Along this process, the growth exponent increases from $\beta\simeq 0.47$ at early times to $\beta\simeq 0.93$ at long times \cite{Santalla2018}, not unlike the type of change we presently obtain for e.g.\ $r_{\rm AB} \simeq 0.1$. The reason for this behavior is the existence of a morphological instability in such type of diffusion-limited systems \cite{Cuerno2001,Nicoli2009}, whereby front protrusions grow faster that front depressions due to their differential exposure to diffusive fluxes. This instability can be triggered along the time evolution of the system, changing the scaling behavior as in Ref.\ [\onlinecite{Santalla2018}]. In the model of patchy colloids, transport is not limited by diffusion. However, the model has an unstable mechanism built in, namely the finite interaction range and colloid alignment step, as suggested by the comparison between the $r_{\rm AB}=1$ case with off-lattice BD. This mechanism is operative for all $r_{\rm AB}$ values and may account for the scaling behavior and macroscopic front shapes (which are not unlike morphologies seen for diffusion-limited systems) that ensue at long times. Qualitatively this agrees with the experiments, in the sense that the morphological instability implied by the colloidal Matthew effect occurs as soon as the colloidal particles are anisotropic with $\varepsilon\neq 1$. Indeed, the supplemental video 3 of the experiments in Ref.\ \cite{Yunker2013} could hardly be told apart from a diffusion-limited aggregation process, see e.g.\ the animation provided with the supplemental material of Ref. \ \cite{Galindo2013}, corresponding to Figure 6 of that reference. We would expect this unstable behavior to persist even in potential 
generalizations of the present model to more realistic conditions for colloidal aggregation systems, such as e.g.\ the case of cluster-cluster aggregation, see e.g.\ Refs.\ \cite{Meakin1998,Jungblut2019} and other therein.

For small $0 \leq r_{\rm AB} \leq 0.01$, the intermediate times (recall Fig.\ \ref{fig:L2048curvasw2}) seem dominated by the scaling obtained for $r_{\rm AB}=0$, that features the largest value of $\beta$ seen in our simulations. Note that the $r_{\rm AB}=0$ behavior is quite similar to that of unstable BD \cite{Asikainen2002b}. The long time behavior for small $r_{\rm AB}>0$ features smaller (but still large) values of $\beta$, which are not far from those seen for large $r_{\rm AB}$ and long times. Such $\beta$ values ensue together with front morphologies which are also dominated by large slopes and macroscopic shapes, again reminiscent of morphologically unstable behavior possibly related with the finite interaction range and colloid attachment rule.

Another important point is the existence of the pinning-depinning transition and its features. In their numerical simulations, Dias {\em et al}.\ \cite{Dias2014,Dias2018} found a non-zero front velocity for all $0 < r_\mathrm{AB} \leq 1$, while at $r_\mathrm{AB}=0$ they argued that the aggregate interface is not well defined. Their reported KPZ-QKPZ crossover at intermediate $r_\mathrm{AB}$ values could be caused by the balance of two competing mechanisms. Indeed, reducing $r_\mathrm{AB}$ hinders binding to B patches. This favors the growth of A-A chains and increases the availability of B sites, which compensates for the reduction by the low $r_\mathrm{AB}$ value. As a result, Dias \textit{et al}.\ argue that the pinning transition does not occur at a single critical $r_{\mathrm{AB},c}$, but within a finite range of $r_\mathrm{AB}$ values. 

This behavior contrasts with that of the QKPZ equation, for which the depinning transition takes place at a point value of the control parameter, and not for a full interval of values, while the average front velocity remains zero right at depinning. In contrast, the use of free boundary conditions in our simulations has allowed us to elucidate a discontinuous pinning transition right at $r_{\rm AB}=0$, while QKPZ scaling exponents are here seen to be effective values. Incidentally, we also note that the occurrence of macroscopic shapes at long times does not depend on the choice of boundary conditions.

\section{Summary and Conclusions}
\label{sect:conclusions}

The kinetic roughening behavior of the coffee-ring aggregates proposed in Ref.~\cite{Dias2018} has been simulated extending their times, sizes and values of $r_\mathrm{AB}$. As a results, we have first characterized a discontinuous pinning-depinning phase transition at $r_\mathrm{AB} = 0$. Particularly, there is a discontinuity in the velocity as $r_\mathrm{AB} \to 0^+$, with $v(r_{\mathrm{AB}} \to 0^+)\neq v(r_\mathrm{AB})=0$. The choice of proper boundary conditions was found to be of paramount importance for $r_\mathrm{AB} = 0$, as only the use of free boundary conditions allows one to elucidate the phase transition. 

We confirm the standard KPZ kinetic roughening behavior reported elsewhere for large $r_\mathrm{AB}$ values and intermediate times. Besides, simulations at large values of $r_\mathrm{AB}$ feature a crossover from 1D-KPZ exponents for intermediate times to large values of the scaling exponents for longer times. However, simulations for smaller values of $r_\mathrm{AB}$ show a strong crossover dominated, at intermediate times, by the $r_\mathrm{AB}=0$ behavior. 

The intermediate (time) dynamics is different depending on the $0 < r_\mathrm{AB} \le 1$ value. For $0<r_\mathrm{AB}\le 0.01$, the system dynamics is similar to that for $r_\mathrm{AB}=0$, the similarity extending to longer times as $r_\mathrm{AB}$ decreases. For these values of $r_\mathrm{AB}$ (including $r_\mathrm{AB} = 0$), we observe  intrinsic anomalous scaling \cite{Lopez1997,Ramasco2000,Cuerno2004} ---a generalization of the simpler Family-Vicsek dynamic scaling Ansatz \cite{Barabasi1995,Krug1997} satisfied e.g.\ by the KPZ equation---, a conclusion which is impossible to reach on the basis of the behavior of the global roughness only. For $r_\mathrm{AB}>0.01$, on the contrary, the intermediate-time behavior is compatible with that of 1D-KPZ, and there is no evidence for intrinsic anomalous scaling. Within this range, exponents are compatible with those of QKPZ, as reported elsewhere. 

For long times, large lattice sizes and all $0 < r_\mathrm{AB}\le 1$ we have found that the system dynamics is dominated by the appearance of macroscopic shapes, whose geometry and growth properties influence the corresponding scaling exponents. Actually, in view of our findings, the critical behavior described by Dias \textit{et al}. ~\cite{Dias2018} (in particular the 1D-QKPZ exponents for their smaller values of $r_\mathrm{AB}$) is an effective behavior influenced by the final macroscopic shapes.

Our simulations expand previous numerical work on this model through the study of correlation functions. This has allowed us to elucidate the occurrence of intrinsic anomalous scaling for all times for small $r_{\rm AB}$, as well as the existence of different scaling behaviors at small and large distances for long times, correlated with the development of macroscopic shapes. Let us note that anomalous scaling is well known to occur for many experimental systems and continuum and discrete models in which morphological instabilities take place \cite{Cuerno2004}, including also cases with faceted interfaces \cite{Ramasco2000}. This fact, combined with the behavior of the scaling exponents and long time front shapes, reinforces our interpretation on the occurrence of a morphological instability dominating the large scale behavior of the patchy colloid model. Comparing with the experiments of Yunker {\em et al}.\ \cite{Yunker2011,Yunker2013}, this was probably to be expected: indeed, the colloidal Matthew effect is a clear morphological instability in which quenched disorder plays no role \cite{Nicoli2013,Yunker2013reply}, and whose relevance to the system behavior increases with the colloid eccentricity. This is precisely the same behavior which the model reproduces for a  decreasing $r_{\rm AB}$.

\begin{acknowledgments}
We are thankful to V\'{\i}ctor Orozco for his participation in a preliminary stage of this project. We also thank C. Dias and M. Telo da Gama for exchanges. This work was partially supported by Ministerio de Ciencia, Innovaci\'on y Universidades (Spain), Agencia Estatal de Investigaci\'on (AEI, Spain), and Fondo Europeo de Desarrollo Regional (FEDER, EU) through Grants PID2020-112936GB-I00, PGC2018-094763-B-I00, and PID2021-123969NB-I00, by the Junta de Extremadura (Spain) and Fondo Europeo de Desarrollo Regional (FEDER, EU) through Grants No.\ GR21014 and IB20079, and by Comunidad de Madrid (Spain) under the Multiannual Agreement with UC3M in the line of Excellence of University Professors (EPUC3M23), in the context of the V Plan Regional de Investigaci\'on Cient\'{\i}fica e Innovaci\'on Tecnol\'ogica (PRICIT). B.\ G.\ Barreales was supported by Junta de Extremadura and Fondo Social Europeo (FSE, EU) through pre-doctoral grant PD18034. We have run our simulations in the computing facilities of the Instituto de Computaci\'{o}n Cient\'{\i}fica Avanzada de Extremadura (ICCAEx).
\end{acknowledgments}

\appendix

\section{Some properties of the quenched KPZ equation}

\label{app:qkpz}

At this point, we briefly review some  key facts on the QKPZ equation; this continuum model reads \cite{Galluccio1995,Stepanow1995}
\begin{align}
    \partial_t h = F+ \nu \nabla^2 &h + \frac{\lambda}{2} (\nabla h)^2 + \eta(\mathbf{r},h) , \label{eq:qkpz} \\
    \langle \eta(\mathbf{r},h) \eta(\mathbf{r}',h') \rangle &= N \, \delta(\mathbf{r}-\mathbf{r}') \delta(h-h') ,
    \nonumber
\end{align}
where $h(\mathbf{r},t)$ is the front position above point $\mathbf{r}$ on a $d$-dimensional substrate, $F$ is a constant external driving force, $\eta(\mathbf{r},h)$ is uncorrelated, zero-average, Gaussian quenched (time-independent) disorder with amplitude $N>0$, and $\nu>0$, $\lambda$ are parameters. This model has a very rich dynamical behavior that has been fully elucidated recently in experiments of reactive fronts in disordered media \cite{Atis2015}, where up to three different universality classes are identified associated with Eq.\ \eqref{eq:qkpz}. First of all, a pinning-depinning transition exists at a non-zero critical value of the driving force $F=F_c$, such that the interface is pinned, i.e.\ the average velocity is zero, or moving (nonzero average velocity) for $F\leq F_c$ or $F>F_c$, respectively \cite{Barabasi1995}. The front described by Eq.\ \eqref{eq:qkpz} only displays the exponent values $\beta\simeq 0.63$ and $z=1$ (associated in Ref.\ \cite{Yunker2013} with the experiments using ellipsoidal colloids of a high eccentricity) exactly at depinning when $F=F_c$ and $\lambda>0$ \cite{Leschhorn1996}. For $F\gg F_c$ and $\lambda>0$, the scaling exponents are those of the standard KPZ universality class \cite{Tang1995,Leschhorn1996}. Finally, for $\lambda <0$, Eq.\ \eqref{eq:qkpz} also describes a pinning transition, but for a very different class (so-called, negative QKPZ) of faceted interfaces \cite{Jeong1996,Jeong1999,Atis2015}. For comparison, the scaling behavior of the KPZ equation [obtained by replacing the disorder $\eta(\mathbf{r},h)$ in Eq.\ \eqref{eq:qkpz} by similarly uncorrelated, time-dependent noise $\eta(\mathbf{r},t)$] depends neither on the value of $F$ nor on the sign of $\lambda$ \cite{Barabasi1995,Krug1997}.

\section{Simulation details} 
\label{app:details}

We study the front propagation in a one-dimensional substrate of size $L$ growing in the perpendicular direction. At $t=0$ the system is empty. At every time step, a particle with radius $R$ falls at a random $x$ position, where it may either bind the substrate or interact with a pre-existing, stationary particle. In particular, if there is a particle in the $(x-\Delta , x+\Delta )$ range, where $\Delta$ is the particle diameter or the width of the columns, the new one interacts with it, or with the highest one of them if there is more than one. For simplicity, $R = 0.5$ and $\Delta =1$ have been chosen without loss of generality. 

When two particles interact, bonding may or may not occur; a schematic of the binding probabilities for our model is shown in Fig.~\ref{fig:schema}. Binding needs first that the interaction ranges of both interacting particles overlap. This range is quantified by the angle $\theta=\pi /6$ around each of the four patches, so that the probability that the interaction ranges of the falling and stationary particles overlap is $4/9$. Secondly, binding depends on the facing patches, the options being AA, AB, BA, and BB. If the pair of facing patches is AA, the bond is always created, since $P(\text{AA})=1$. In all other cases, the bond is created with a probability $r_\mathrm{AB}$ which is a system parameter: $P(\text{AB})=P(\text{BA})=P(\text{BB})=r_\mathrm{AB}$. When the falling particle binds, its patches line up with those of the pre-existing particle to which it becomes attached. Time increases by one unit each time a new particle falls, whether there is binding or not.

\begin{table}[h]
\schema
{
	\schemabox{\hspace{0.1cm}}
}
{
	\schemabox{1/3 \\ Falling particle: \\ Oriented outside \\ the range of interaction. \\ There is no binding.\\ \\
		\schema
		{
			\schemabox{2/3 \\ Falling particle: \\ A or B oriented.}
		}
		{
			\schemabox{1/3 \\  Stationary particle: \\ Oriented outside \\ the range of interaction. \\ There is no binding. \\  \\ 
			\schema
			{
			    \schemabox{2/3 \\ Stationary particle: \\ A or B oriented.}
			}
			{
			    \schemabox{1/4 \\ AA bond \\ \\ $3~ r_\mathrm{AB}/4$ \\ AB, BA or \\ BB bond \\ \\ 3~($1-r_\mathrm{AB}$)/4 \\ There is \\ no binding.}
			}}}
		}
}
\renewcommand{\tablename}{}
    \caption*{Schema of bond formation.}
    \label{fig:schema}
\end{table}

Parameter conditions for the entire set of runs reported herein are collected in Table~\ref{tab:details}.

\begin{table}[!t]
\caption{\label{tab:details}
Parameter values for our numerical simulations.}
\begin{ruledtabular}
\begin{tabular}{cdrc}
 $L$ & \multicolumn{1}{c}{$r_\mathrm{AB}$}  & \multicolumn{1}{c}{$t_{\text{max}}$} & runs \\
\hline
\multirow{10}{*}{128}  & 1 & 100$\times$10$^3$ & 200 \\
& 0.5 & 100$\times$10$^3$ & 200 \\
& 0.4 & 100$\times$10$^3$ & 200 \\
& 0.3 & 100$\times$10$^3$ & 200 \\
& 0.2 & 100$\times$10$^3$ & 200 \\
& 0.1 & 100$\times$10$^3$ & 200 \\
& 0.01 & 100$\times$10$^3$ & 200 \\
& 0.001 & 100$\times$10$^3$ & 200 \\
& 0.0001 & 100$\times$10$^3$ & 200 \\
& 0 & 200$\times$10$^3$ & 200 \\ \hline
\multirow{10}{*}{512}  & 1 & 30$\times$10$^3$ & 200 \\
& 0.5 & 42$\times$10$^3$ & 200 \\
& 0.4 & 46$\times$10$^3$ & 200 \\
& 0.3 & 50$\times$10$^3$ & 200 \\
& 0.2 & 57$\times$10$^3$ & 200 \\
& 0.1 & 82$\times$10$^3$ & 200 \\
& 0.01 & 100$\times$10$^3$ & 200 \\
& 0.001 & 110$\times$10$^3$ & 200 \\
& 0.0001 & 300$\times$10$^3$ & 200 \\
& 0 & 400$\times$10$^3$ & 200 \\ \hline
\multirow{10}{*}{2048}  & 1 & 20$\times$10$^3$ & 40 \\
& 0.5 & 20$\times$10$^3$ & 40 \\
& 0.4 & 20$\times$10$^3$ & 40 \\
& 0.3 & 20$\times$10$^3$ & 40 \\
& 0.2 & 20$\times$10$^3$ & 40 \\
& 0.1 & 30$\times$10$^3$ & 40 \\
& 0.01 & 70$\times$10$^3$ & 40 \\
& 0.001 & 100$\times$10$^3$ & 40 \\
& 0.0001 & 120$\times$10$^3$ & 40 \\
& 0 & 130$\times$10$^3$ & 40 \\
& \mathrm{BD} & 20$\times$10$^3$ & 420 \\
\end{tabular}
\end{ruledtabular}
\end{table}

\section{Free versus periodic boundary conditions} \label{app:PBC}

\begin{figure*}
    \centering
    \includegraphics[width=0.9\textwidth]{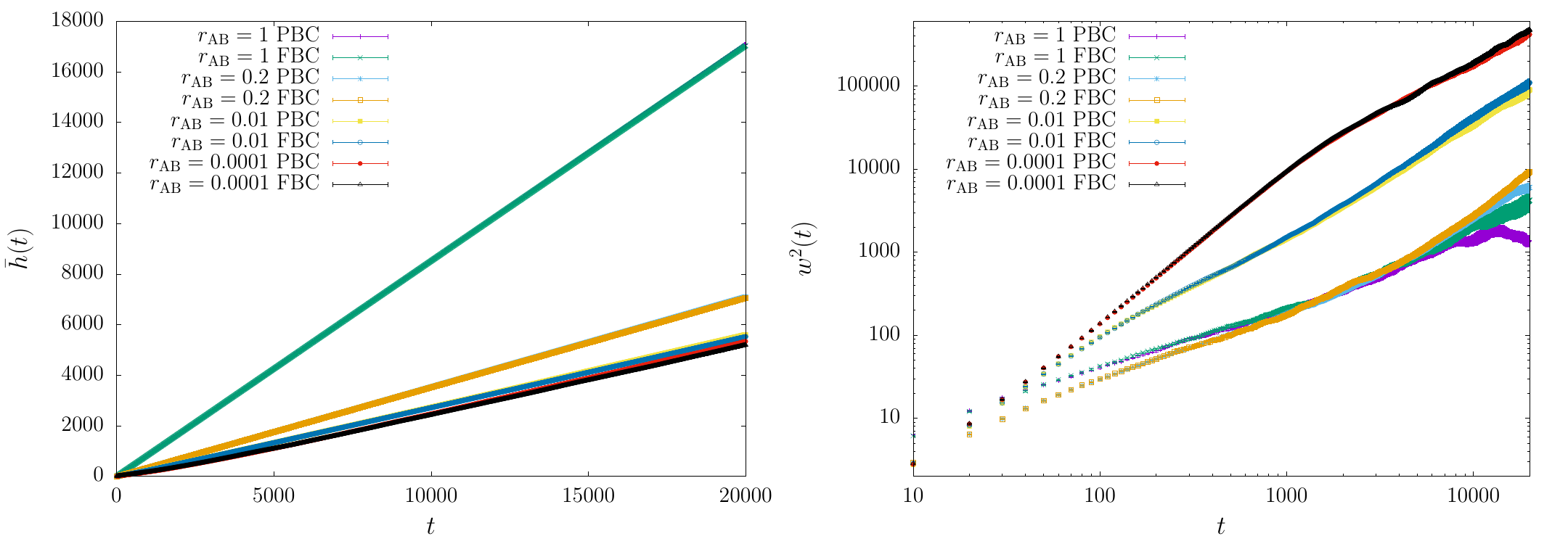}
    \caption{Mean front height (left panel) and squared roughness (right panel) as functions of time for values of $p=r_\mathrm{AB}$ as given in each legend, using FBC or PBC as indicated. For some sets of data, FBC data hide total or partially PBC results. All units are arbitrary.}
    \label{fig:pbc_vel_w2}
\end{figure*}

To check the influence of the boundary conditions on the overall dynamic behavior of the system, we have computed a number of quantities for several values of $r_\mathrm{AB}$ in the same conditions ($L=2048$ and 20 runs) using FBC and PBC. The mean front and the squared roughness as functions of time are shown in the left and right panels of Fig.\ \ref{fig:pbc_vel_w2}, respectively. The front evolution is virtually identical for PBC and FBC, within error bars. The roughness curves do show some differences at long times, probably due to the larger front fluctuations at saturation in these conditions. In any case, results reported in the main text are well below such time intervals, so that no effect of the boundary conditions is to be expected in our results.

\section{Off-lattice ballistic deposition} \label{app:BD}

As noted in Sec.\ \ref{sect:TheModel}, one might naturally expect the model of Refs.\ \cite{Dias2014,Dias2018} to behave as simple off-lattice ballistic deposition \cite{Meakin1998} in the limiting $r_{\rm AB}=1$ case, as the difference is lost between the $A$ and $B$ poles. However, the finite interaction zone and the alignment of the particles after attachment both persist, which do not occur in the off-lattice ballistic deposition model. We have already seen in Fig.\ \ref{fig:L2048curvasw2} that this difference suffices to change the behavior of the front roughness of the patchy colloid model for $r_{\rm AB}=1$ with respect to that of off-lattice BD at long times. In this Appendix we collect further results from numerical simulations that we have performed of off-lattice BD, that can readily be contrasted with those discussed in the main text for the model of patchy colloidal particles.

\begin{figure}[htb]
    \centering
    \includegraphics[width=0.45\textwidth]{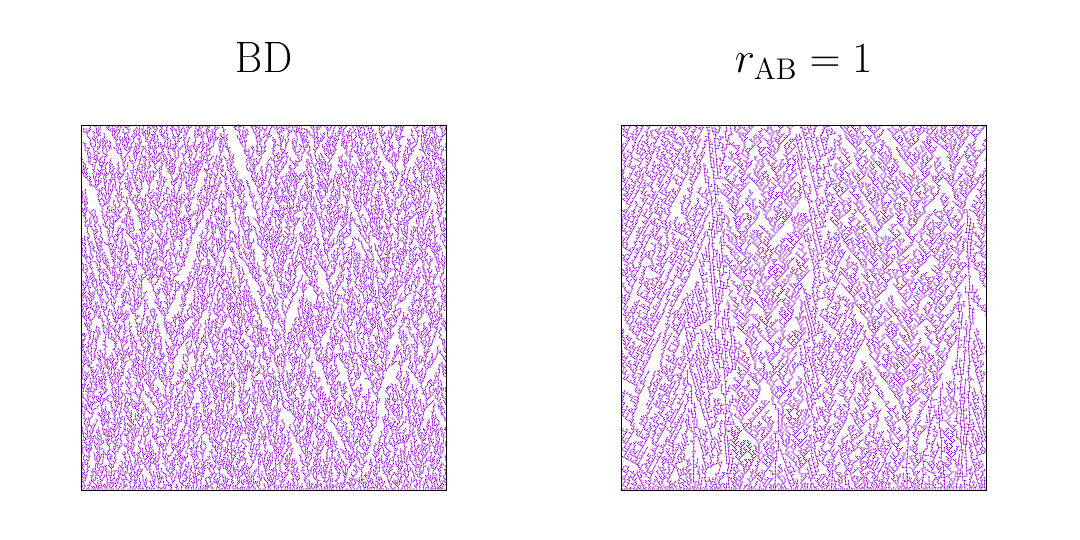}
     \caption{Morphologies of the colloidal aggregates for off-lattice ballistic deposition and patchy colloids model $r_{\rm AB}=1$ case. The size of the simulations is $L=2048$ but we show only the range [0:255] for easier view. Times ranging from the initial ones to final times for which the points of the morphologies fill the plots. All units are arbitrary.}
    \label{fig:BD_morf_inicial}
\end{figure}

Figure \ref{fig:BD_morf_inicial} shows colloids aggregates for off-lattice ballistic deposition and for patchy colloids model in the case $r_{\rm AB}=1$, at the beginning of two runs. We can appreciate subtle differences, such as that in the case of $r_{\rm AB}=1$ there are allowed directions that create perpendicular branches and the system is less dense. 

\begin{figure}[htb]
    \centering
    \includegraphics[width=0.45\textwidth]{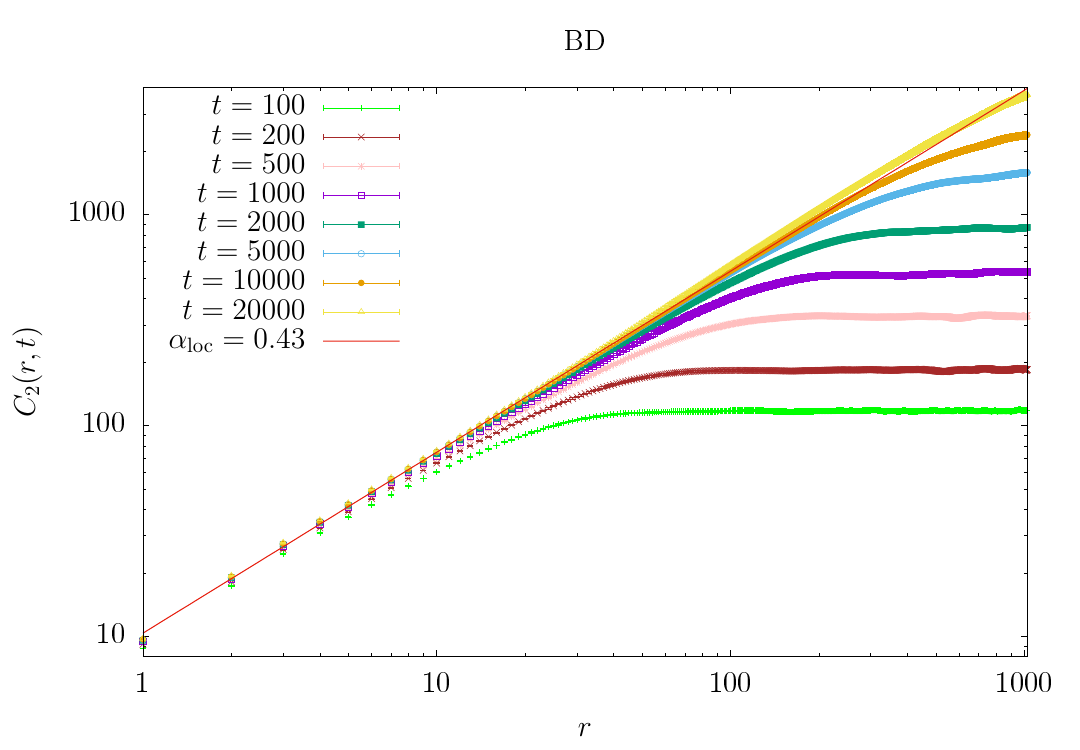}
     \caption{Front correlation function $C_2(r,t)$ vs $r$ for off-lattice BD. Solid line corresponds to fit to the power law $r^{2\alpha_{\mathrm{loc}}}$ for $t=20000$ and $r<200$. All units are arbitrary.}
    \label{fig:BD_c2}
\end{figure}

\begin{figure}[htb]
    \centering
    \includegraphics[width=0.45\textwidth]{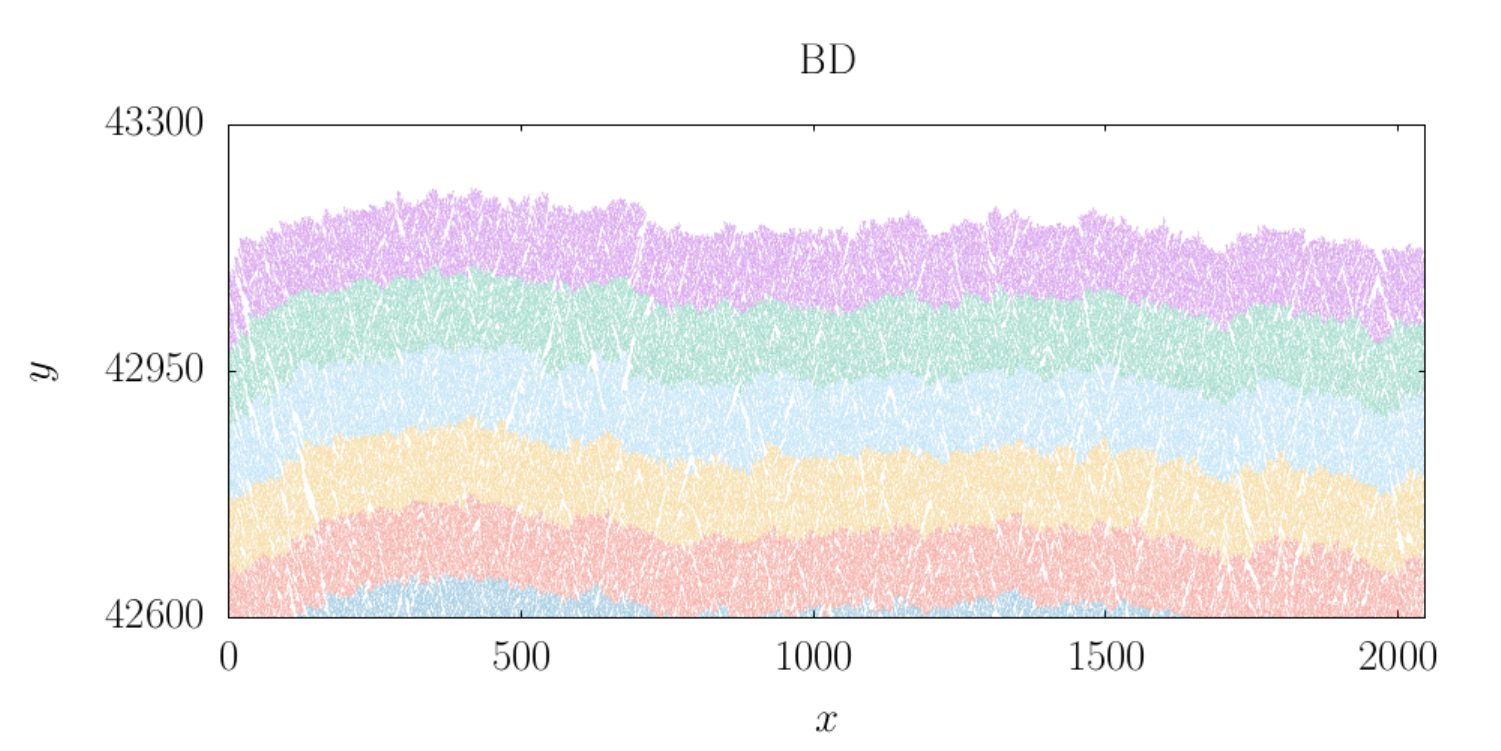}
     \caption{Morphology of a colloidal aggregate at the final time, $t=20000$, for off-lattice BD.  Each color shows the last hundred thousand particles to join the system. All units are arbitrary.}
    \label{fig:BD_morf_final}
\end{figure}

As mentioned in section \ref{sec:correlation_function}, the height correlation function for off-lattice BD (see Fig.~\ref{fig:BD_c2}) shows a standard scaling behavior which contrasts with those shown in Fig.~\ref{fig:c2} for the patchy colloids model. Consistent with this, the morphologies at long times, $t=20000$, for off-lattice BD show a height front smoother compared to $r_{\rm AB}=1$ case, as can be seen in Fig.~\ref{fig:BD_morf_final} and in Fig.~\ref{fig:morf_final}. For completeness, the scaling exponents computed for the off-lattice BD model are $z =$ 1.41(2) and $\alpha =$ 0.469(7), consistent with those reported elsewhere \cite{Meakin1998}.

\end{document}